\documentclass{ws-ijmpb}
\usepackage{color}

\newcommand{\avg}[1]{\left\langle {#1} \right\rangle}

\newcommand{\beq}{\begin{equation}}
\newcommand{\eeq}{\end{equation}}
\newcommand{\bea}{\begin{eqnarray}}
\newcommand{\eea}{\end{eqnarray}}

\renewcommand{\d}{{\rm d}}

\begin{document}

\markboth{R.~Hamerly et al.}
{Topological defect formation in 1D and 2D spin chains realized by network of OPOs}

%
\catchline{}{}{}{}{}
%

\title{Topological defect formation in 1D and 2D spin chains realized by network of optical parametric oscillators}


\author{
	Ryan Hamerly${}^1$\footnote{\email{rhamerly@stanford.edu}}\ ,
	Kensuke Inaba${}^2$,
	Takahiro Inagaki${}^2$,
	Hiroki Takesue${}^2$,\\
	Yoshihisa Yamamoto${}^{1,3}$
	and Hideo Mabuchi${}^1$}
	
\address{
	${}^1$Edward L.\ Ginzton Laboratory, Stanford University, Stanford, CA 94305\\
	${}^2$NTT Basic Research Laboratories, NTT Corporation, 3-1 Morinosato Wakamiya, Atsugi, Kanagawa, 243-0198, Japan\\
	${}^3$ImPACT Program, Japan Science and Technology Agency, 7 Gobancho, Chiyoda-ku, Tokyo 102-0076, Japan}

\maketitle

\begin{history}
\received{May 26, 2016}
\end{history}

\begin{abstract}
A network of optical parametric oscillators is used to simulate classical Ising and XY spin chains.  The collective nonlinear dynamics of this network, driven by quantum noise rather than thermal fluctuations, seeks out the Ising / XY ground state as the system transitions from below to above the lasing threshold.  We study the behavior of this ``Ising machine'' for three canonical problems: a 1D ferromagnetic spin chain, a 2D square lattice, and problems where next-nearest-neighbor couplings give rise to frustration.  If the pump turn-on time is finite, topological defects form (domain walls for the Ising model, winding number and vortices for XY) and their density can be predicted from a numerical model involving a linear ``growth stage'' and a nonlinear ``saturation stage''.  These predictions are compared against recent data for a 10,000-spin 1D Ising machine.
\end{abstract}


\section{Introduction}

Many important problems in computer science can be solved by message-passing algorithms.  In such algorithms, information lives on the nodes of a graph, while computation consists of updating the values of the nodes by passing ``messages'' along the graph's edges.  Examples of such algorithms include neural networks\cite{Izhikevich2007}, probabilistic graphical models\cite{Koller2009}, low-density parity check codes\cite{Pavlichin2014} and topological surface codes\cite{Fujii2014}.  Message-passing algorithms are advantageous because they are intrinsically parallel, making them straightforward to implement on multi-core architectures.

As digital microprocessors reach their physical limits, there has been a surge of research into special-purpose hardware for various message-passing algorithms.  In electronics, examples include CMOS artificial neural networks\cite{Benjamin2014,Merolla2014,Misra2010,Schemmel2010} and CMOS chips for simulated-annealing\cite{Yoshimura2015}.  Quantum annealers have a similar graphical architecture, with data stored at the vertices (qubits), while pairwise couplings along the edges transmit information along the graph.

This paper focuses on a coherent {\it optical} network, which functions as a message-passing algorithm to solve the {\it Ising problem} and the related {\it XY problem}.  These problems consist of finding the global minimum of the Ising potential $\min_\sigma[U(\sigma)]$, where
\beq
	U(\sigma) = -\frac{1}{2}\sum_{ij} {J_{ij} \vec{\sigma}_i \cdot \vec{\sigma}_j} \label{eq:09-u}
\eeq
In (\ref{eq:09-u}), $J_{ij}$ is the coupling between spins $\vec{\sigma}_i$, $\vec{\sigma}_j$.  The spins $\vec{\sigma} \in \mathbb{R}^d$ have unit norm $|\vec{\sigma}| = 1$.  For the Ising problem $d = 1$ and $\vec{\sigma} \in \mathbb{Z}_2 = \{-1, 1\}$; for the XY problem $d = 2$ and $\vec{\sigma} \in S_1 = U(1)$.  Higher-dimensional problems ($d = 3,4,5\ldots$) can also be defined, but will not be considered here.

The general Ising problem is NP-hard\cite{Barahona1982}, but algorithms based on convex relaxation or heuristics can give approximate solutions in polynomial time.  A number of schemes have been studied to map such algorithms directly onto electronic\cite{Yoshimura2015} or photonic circuits\cite{Hamerly2015-2}.

\begin{figure}[tbp]
\begin{center}
\includegraphics[width=1.0\textwidth]{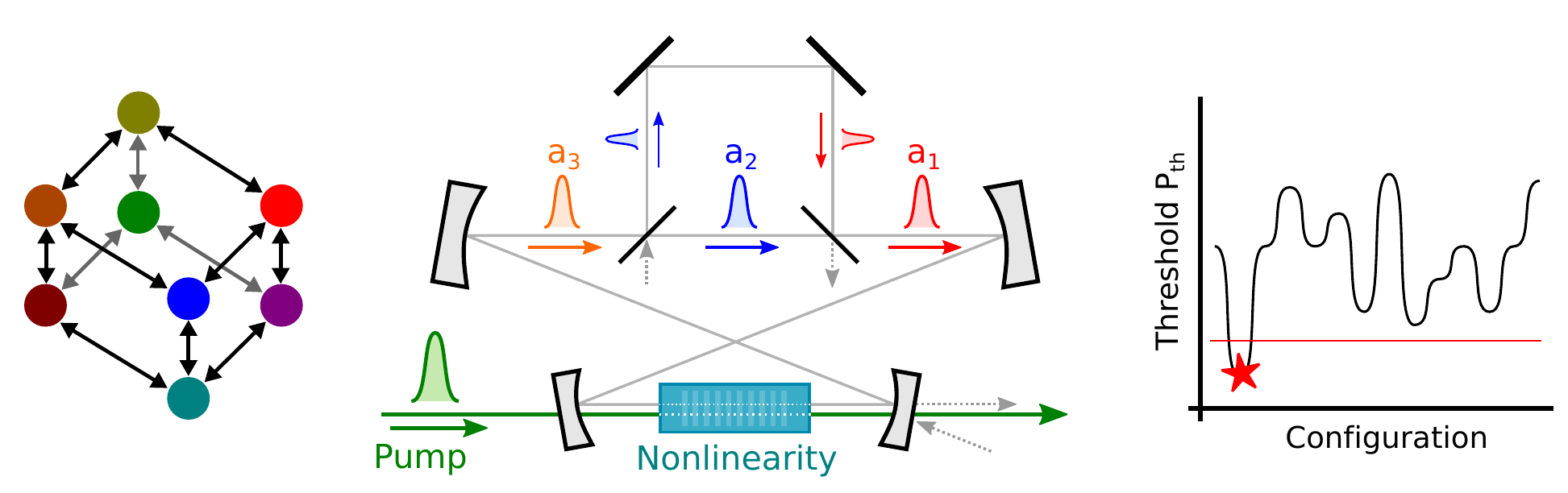}
\caption{Left: Ising machine consisting of optical gain elements (circles) with injection coupling (arrows), cubic graph.  Center: time-multiplexed implementation.  Right: illustration of the minimum-gain principle.}
\label{fig:09-f1}
\end{center}
\end{figure}

The coherent Ising machine is a network of identical nonlinear gain elements symmetrically coupled by optical injection that solves the Ising problem by a minimum-gain principle\cite{Haribara2016,Wang2013,Marandi2014}.  According to this principle, if the couplings are chosen to implement the potential $J_{ij}$, the configuration that oscillates should minimize the potential (\ref{eq:09-u}).  For the nonlinear gain, an injection-locked laser or an optical parametric oscillator (OPO) can be used.  In practice, the spins in the machine are time-multiplexed as pulses in a synchronously-pumped laser or OPO and couplings are realized by delay lines that couple pulses at different locations in the cavity (Fig.~\ref{fig:09-f1}).

The Ising machine was proposed as an injection-locked laser network\cite{Takata2012,Utsunomiya2011}.  Later, the theory was extended to OPOs\cite{Wang2013} and simulations showed promising performance on MAX-CUT Ising problems of size $N \leq 20$.  Experimental results followed for an $N=4$ OPO network\cite{Marandi2014} and an $N = 16$ network\cite{KentaThesis,Takata2016}, as well as simulations for G-set graphs\cite{Haribara2015} up to $N = 20000$.

In this paper, we analyze the OPO Ising machine for solving the simplest class of Ising problems: 1D and 2D ferromagnetic chains.  Although these problems are trivial in the sense that the solutions are well-known, the analytic theory one can derive gives the reader a more lucid understanding of how the Ising machine actually works.  Because of their simplicity, 1D and 2D models may serve as a good way to ``benchmark'' the performance of different Ising machines.  Moreover, they are one of the simplest systems to realize in the laboratory, requiring only one delay line, allowing for direct comparison between the theory and currently realizable experiments.  As a model experimental system, we use the four-wave mixing fiber OPO implemented in our previous paper\cite{Inagaki2016}.

Section \ref{sec:theory} covers the theory of the time-multiplexed OPO Ising machine.  Based on this theory, we derive semiclassical equations of motion for the OPO pulse amplitudes.  In the original formulation of the Ising machine as a network of continuous-wave OPOs, these are stochastic differential equations\cite{Wang2013}, but for the pulsed case we show that they become {\it difference equations}, relating the pulse amplitudes between successive round trips.

These equations are solved in Sec.~\ref{sec:09-coll}, where we show that the dynamics breaks down into two stages: a {\it growth stage} where the field amplitudes are well below threshold and growth is linear, and a {\it saturation stage} where the OPO amplitudes saturate, giving rise to nonlinear dynamics defined by domains and domain walls.  Using this picture, in Sec.~\ref{sec:statistics} we derive expressions for the correlation length, domain-wall density and domain-length histogram for Ising machine solution.  This is compared to experimental data from the fiber OPO of Inagaki et al.\cite{Inagaki2016}; we show that our theory matches the experimental results, while a simple thermal Ising model does not.

Sections \ref{sec:09-2d}-\ref{sec:xy1d2d} explore more complex systems that have not yet been realized in OPO experiments.  In Sec.~\ref{sec:09-2d}, the two-dimensional lattice is treated.  The same growth / saturation stage picture applies, but during the latter we find 2D domains separated by 1D domain walls which move towards their center of curvature and collapse in a time quadratic in the domain size.  XY models are treated in Sec.~\ref{sec:xy}-\ref{sec:xy1d2d}, where the basic equations are introduced and applied to 1D and 2D systems.  Instead of domains, the XY model gives winding-number states for the 1D chain and vortices for 2D.  These vortices resemble those from Berezinskii-Kosterlitz-Thouless theory\cite{Berezinskii1971,Kosterlitz1974,Kosterlitz1973}, but they are generated by a non-thermal mechanism, and so their distribution is also athermal.

\section{Fiber OPO Theory}
\label{sec:theory}

First, we derive equations of motion for the pulse amplitudes in the cavity.  For concreteness, we consider the case of a singly resonant $\chi^{(3)}$ fiber OPO (typical parameters, following Inagaki et al.\cite{Inagaki2016} are given in Table \ref{tab:09-t1}).  In this system, a narrowband filter ensures that the signal $a_i(t)$ is resonant, while the pump fields $b_i(t)$, $c_i(t)$ are not.  The nonlinearity is provided by the degenerate four-wave mixing process $2\omega_a \leftrightarrow \omega_b + \omega_c$ in the nonlinear fiber.

\begin{table}[tb]
\tbl{Typical parameters for a pulsed four-wave mixing fiber OPO Ising machine.}
{\begin{tabular}{c|cc}
\Hline
Term & Value & Description \\
\hline 
$\lambda_a$ & 1541 nm & Signal wavelength \\
$\lambda_b, \lambda_c$ & 1552 nm, 1531 nm & Pump wavelengths \\
$\gamma$ & $21\;\mbox{W}^{-1}\mbox{km}^{-1}$ & Fiber nonlinearity \\
$G_0$ & 7 dB & Fiber gain at threshold \\
$r, t$ & $1/\sqrt{2}$ & Delay mirror coefficients, $r^2 + t^2 = 1$ \\
$f$ & 2 GHz & Pulse frequency (time between pulses is $1/f$) \\
$\tau$ & 60 ps & Pulse width \\
$N$ & 10000 & Number of pulses \\ \Hline
\end{tabular}}
\label{tab:09-t1}
\end{table}

If the OPO network is viewed as a computer, the ``memory'' is stored in the signal pulse amplitudes $a_i(t)$, $i \in \{0, 1, 2, \ldots, N-1\}$ is the pulse index and $t \in \{0, 1, 2, \ldots\}$ is the round-trip number, which serves as a discretized time.  The ``processor'' consists of the $\chi^{(3)}$ fiber, a nonlinear map which acts on each pulse independently; and the delay line(s), which create a linear coupling between the pulses.  The ``inputs'' are the amplitudes of the pump pulses $b_i(t)$, $c_i(t)$, which can be programmed with an amplitude modulator placed in front of the pump laser.

Each round trip can be modeled as a cascade of three operations: nonlinear gain, coupling, and linear loss.  Ignoring vacuum noise, this gives the following map:
\beq
	a_i(t) \stackrel{\rm Fiber}{\longrightarrow} \frac{F[a_i(t)]}{e^{\alpha L_{\rm eff}/2}}
	\stackrel{\rm Loss}{\longrightarrow} \frac{F[a_i(t)]}{\sqrt{G_0}} 
	\stackrel{\rm Coupling}{\longrightarrow} \underbrace{\sum_j C_{ij}\frac{F[a_j(t)]}{\sqrt{G_0}}}_{a_i(t+1)} \label{eq:09-diffeq}
\eeq
Equation (\ref{eq:09-diffeq}) relates $a_i(t+1)$ to $a_i(t)$, giving us an equation of motion for the OPO network.  In the sections below, we obtain the nonlinear gain function $F[a_i(t)]$ and the coupling matrix $C_{ij}$, that form the core of (\ref{eq:09-diffeq}).  Once these are known, Ising machines of arbitrary complexity can be simulated.

\subsection{Nonlinear Fiber}

In the highly nonlinear fiber, the $\chi^{(3)}$ term gives rise to self-phase modulation (SPM) cross-phase modulation (XPM), and degenerate four-wave mixing (DFWM).  In the limit $|b|^2 \ll |c|^2$ with $c$ a flat-top pulse, SPM and XPM give constant phase shifts and can be cancelled by the appropriate phase matching\cite{AgrawalBook}, leaving only the DFWM term.  In this paper we assume that the pulses are sufficiently long that the pulse amplitude is a constant (in time) and dispersion can be neglected; in this case the fields $a, b, c$ depend only on the distance $z$ the fiber, and the fiber field equations are\cite{AgrawalBook,BoydBook}:

\bea
	\frac{\d a}{\d z} & = & \gamma a^* b c - \frac{1}{2}\alpha a \\
	\frac{\d b}{\d z} & = & -\frac{1}{2}\gamma a^2 c^* - \frac{1}{2}\alpha b \\
	\frac{\d c}{\d z} & = & -\frac{1}{2}\gamma a^2 b^* - \frac{1}{2}\alpha c		
\eea

One can rescale the dependent variables $(a, b, c)$ to eliminate the constant $\gamma$; likewise, one can transform the independent variable $t$ to get rid of the linear absorption term.  With the field rescaling $x = (\gamma L_{\rm eff})^{-1/2}e^{-\alpha z/2} \bar{x}$ ($x = a, b, c$, $L_{\rm eff} = (1-e^{-\alpha L})/\alpha$) and length scaling $s = (1 - e^{-\alpha z})/(1 - e^{-\alpha L})$, the equations simplify to
\beq
	\frac{\d \bar{a}}{\d s} = \bar{a}^*\bar{b}\bar{c},\ \ \ 
	\frac{\d \bar{b}}{\d s} = -\frac{1}{2} \bar{a}^2 \bar{c}^*,\ \ \ 
	\frac{\d \bar{c}}{\d s} = -\frac{1}{2} \bar{a}^2 \bar{b}^* \label{eq:feom}
\eeq
and are solved on the interval $s \in [0, 1]$.  Gain occurs when $\bar{a}, \bar{b}, \bar{c}$ satisfy the correct phase relation.  Up to a global phase shift, this requires that $\bar{a}, \bar{b}, \bar{c}$ all be real and positive.  Taking $c(s) = c_{\rm in}$ constant since $c \gg a, b$, and using the constant of motion $B = \bar{b}^2 + \bar{a}^2/2$ from detailed balance, one derives:
\beq
	\frac{\d \bar{a}}{\d s} = c_{\rm in} \bar{a} \sqrt{B^2 - \bar{a}^2/2} \label{eq:09-dads}
\eeq

The fiber output is $a(z=L) = e^{-\alpha L_{\rm eff}/2} \bar{a}(s=1)$.  If the total cavity loss is $G_0$, then the field passes through an additional loss term $\sqrt{G_0 e^{\alpha L_{\rm eff}}}$, giving $a_{\rm out} = G_0^{-1/2} \bar{a}(1)$.  We find
\beq
	\bar{a}(1) = \bar{a}(0) e^{B \bar{c}_{\rm in}} \left[1 + (e^{2B\bar{c}_{\rm in}} - 1)\frac{1 - \sqrt{1 - \bar{a}(0)^2/2B^2}}{2}\right]^{-1} \label{eq:09-a1}
\eeq

The strong pump $c$ has a fixed amplitude, while the weak pump $b$ can be varied.  Define $b_0$ as the cavity threshold in the absence of coupling.  Linearizing (\ref{eq:09-a1}) in the limit $a \ll b$, we find that threshold is achieved when $e^{\bar{b}_0 \bar{c}_{\rm in}} = G_0^{1/2}$.  In terms of the $G_0$, the fiber input-output relation including both gain and loss is:
\beq
	a_{\rm out} = a_{\rm in} G_0^{\frac{1}{2}\bigl(\sqrt{(b_{\rm in}^2 + a_{\rm in}^2/2)}/b_0 - 1\bigr)}
	\left[1 + \bigl(G_0^{\sqrt{(b_{\rm in}^2 + a_{\rm in}^2/2)}/b_0} - 1\bigr) \frac{1 - \sqrt{2b_{\rm in}^2/(a_{\rm in}^2 + 2b_{\rm in}^2)}}{2}\right]^{-1} \label{eq:09-aout}
\eeq
Equation (\ref{eq:09-aout}) has two limits.  When $a_{\rm in} \ll b_{\rm in}$, the terms in the square brackets can be ignored and the field experiences linear gain: $a_{\rm out} = G_0^{\frac{1}{2}(b/b_0 - 1)} a_{\rm in}$.  Thus, the (power) gain for the fiber above threshold is $G_0^{b/b_0}$, and when $b > b_0$ this exceeds the cavity loss.  On the other hand, when $a_{\rm in} \gg b_{\rm in}$, the exponential inside the square brackets dominates and the field is substantially reduced.  This is the DFWM process working in reverse.

From quantum mechanics we know that the field is not defined by a scalar variable $a_i(t)$ but by a state in a harmonic potential.  In the truncated Wigner picture this gives rise to vacuum noise in the signal and pump fields\cite{Carter1995,Kinsler1991,Santori2014}.  To treat this, we need to add fluctuations to the fields $b, c$ before they are inserted: $b_{i,\rm in} \rightarrow b_{i,\rm in} + w^{(b)}_i$, $c_{i,\rm in} \rightarrow c_{i,\rm in} + w^{(c)}_i$, where $w^{(b,c)}_i$ are complex Gaussians that satisfy $\langle w^* w\rangle = \tfrac{1}{2}$, $\langle w \rangle = \langle w^2 \rangle = 0$.  This is the discrete-time analogue of vacuum noise.

To account for the quantum noise in $a_i$, it is easiest to assume that the loss happens in a lumped element after the fiber, rather than concurrently with the gain.  Near threshold, this is a reasonable approximation; elsewhere the noise is larger by a constant $O(1)$ factor.  Making use of this assumption, one must add vacuum fluctuations $a_{i,\rm out} \rightarrow a_{i,\rm out} + \sqrt{1 - 1/G_0} w^{(a)}_i$ to the signal.

All of these results can be applied to $\chi^{(2)}$ OPOs because the strong pump $c$ was presumed constant.  Removing it from Eqs.~(\ref{eq:feom}), one recovers the standard SHG equations, with $\epsilon = \gamma c$ as the $\chi^{(2)}$ parameter.

For a more realistic treatment of the pulsed OPO, one must abandon the continuous-wave picture in Eqs.~(\ref{eq:feom}) and treat the pulse shape itself as a dynamical variable.  The result is a ``multimode'' theory of the OPO, where the actual pulse is a weighted sum of normal modes.  This is a topic unto itself, which we have treated at length in a separate paper\cite{Hamerly2016MM}; a key finding is that if the cavity dispersion is large enough, or a sufficiently narrowband filter is inserted in the cavity, only a single normal mode resonates, and multimode effects can be ignored.  Although the multimode theory changes the exact expression $F(a_{\rm in})$, this ultimately does not matter.  We show in subsequent sections that the performance of the Ising machine depends only on the general form of $F(a_{\rm in})$: the gain at threshold and the near-threshold saturation (which goes as $O(a_{\rm in}^3)$).

\subsection{Coupling}

\begin{figure}[b!]
\begin{center}
\includegraphics[width=0.9\textwidth]{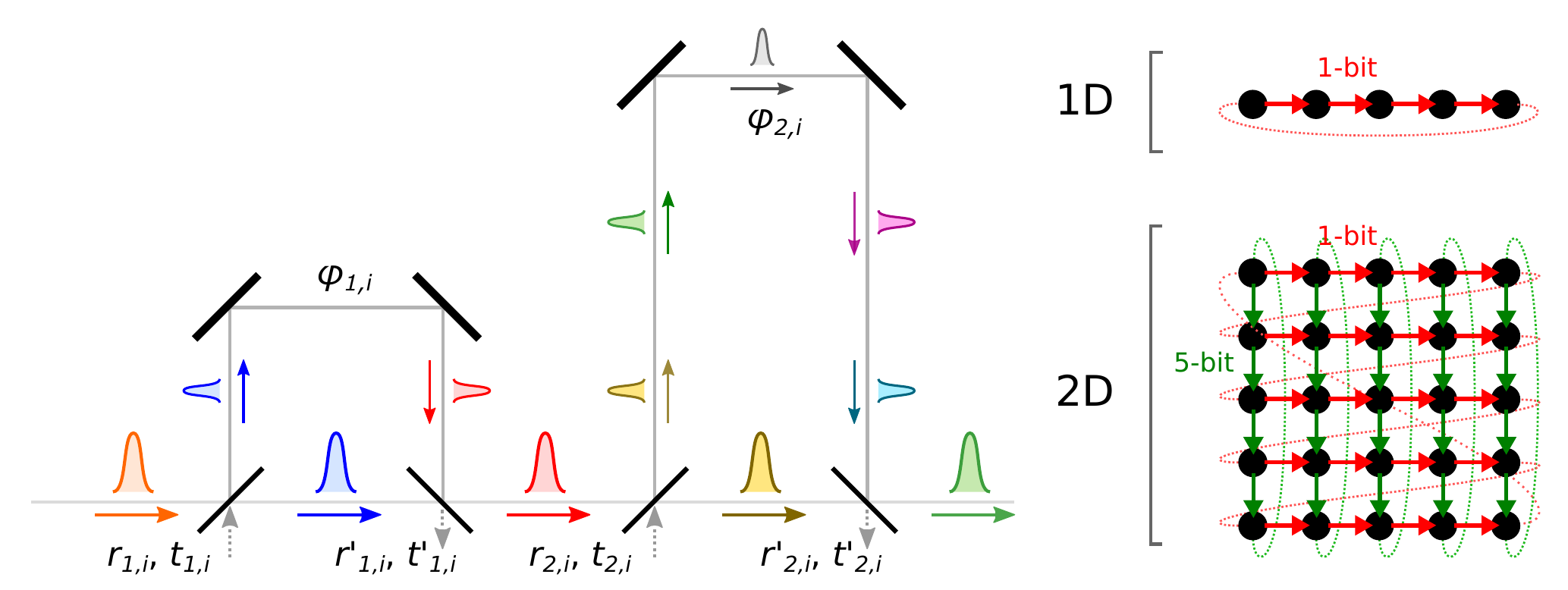}
\caption{Left: schematic a 1-bit and 5-bit delay line.  Right: Ising graphs implemented $N = 5$ with 1-bit delay (top), $N=25$ with 1-bit and 5-bit delay (bottom).}
\label{fig:09-f2}
\end{center}
\end{figure}

This section considers inter-pulse couplings mediated by delay lines and beamsplitters (Fig.~\ref{fig:09-f2}).  Recent experiments all use delay-line couplings\cite{Inagaki2016,KentaThesis,Marandi2014}, although it poses difficulties when many delay lines are involved.  A $d$-bit delay has five parameters: $r, t, r', t', \phi$, where $r^2 + t^2 = 1$, $(r')^2 + (t')^2 = 1$.  With fast modulators, in principle one can make all of these parameters (except $d$) pulse-dependent, giving them an index $i$.  Tracing the paths in Figure \ref{fig:09-f2}, and including the vacuum that enters through the lower-left beamsplitter, the input-output relation for a single delay is:
\beq
	a_i \rightarrow t'_i t_i a_i + r'_i r_{i-d} e^{i\phi_i} a_{i-d} + \left(t'_i r_i w^{(J)}_i + r'_i t_{i-d} e^{i\phi}w^{(J)}_{i-d}\right) \label{eq:09-ai-delay}
\eeq
where the $w_i^{(J)}$ are vacuum processes with $\langle w^* w\rangle = \tfrac{1}{2}$.  One must be careful to avoid negative indices: for instance $a_{-1}(t)$ maps to $a_{N-1}(t-1)$.

If the delays are static and $t = t'$, then (\ref{eq:09-ai-delay}) takes the simplified form:
\beq
	a_i \rightarrow t^2 a_i + r^2 e^{i\phi} a_{i-d} + tr \left(w^{(J)}_i + e^{i\phi}w^{(J)}_{i-d}\right) \label{eq:09-ai-delay2}
\eeq
This section will focus on the static-delay limit, since the experiments to date use static delays.  But the theory and code can accommodate the arbitrary case.

By relabeling the paths so that the long path is the ``cavity'' path and the short path is the ``delay'', and swapping $r \leftrightarrow t$, a delay can be converted into an ``advance'', which mixes $a_i$ with $a_{i+d}$ (again one must be careful with labeling; $a_N(t)$ corresponds to $a_0(t+1)$).  The cavity is enlarged by $d$, so $N \rightarrow N + d$.  Thus it is possible to engineer symmetric length-$d$ couplings using two identical $d$-bit delays.

A 1-bit delay implements the nearest-neighbor coupling of a 1D Ising chain.  To implement a 2D $m \times n$ lattice, one needs a 1-bit delay for the horizontal coupling and an $m$-bit delay for the vertical.  This gives a lattice with periodic but ``offset'' boundary conditions, as shown in Figure \ref{fig:09-f2}.  To implement the lattice without the offsets requires three delays, with time dependence; that case is not treated here.

\subsection{Linear and Near-Threshold Limits}
\label{sec:09-limits}

The fiber OPO has two analytically tractable limits: the linear case $a \ll b$ and the near-threshold case $b \approx b_0$.  These limits arise when we expand the fiber input-output relation (\ref{eq:09-aout}) to third order in $a_{\rm in}$:
\beq
	a_{\rm out} = a_{\rm in} \sqrt{G/G_0} \left[1 - \frac{G - (1 + \log G)}{8} (a_{\rm in}/b)^2 + O\left((a_{\rm in}/b)^4\right)\right] \label{eq:09-inout3}
\eeq
where
\beq
	G = G_0^{b/b_0} \label{eq:09-gain}
\eeq
The {\it linear limit} applies when $a \ll b$.  Taking only the linear term in (\ref{eq:09-inout3}) and combining it with (\ref{eq:09-ai-delay2}), one finds (for a single $d$-bit delay):
\beq
	a_i(t+1) = G_0^{\tfrac{1}{2}(b/b_0 - 1)} \left[t^2 a_i + r^2 a_{i-d}\right] + \mbox{(noise terms)} \label{eq:09-linear}
\eeq
The {\it near-threshold limit} applies when $b \approx b_0$.  In this case, (\ref{eq:09-inout3}) is expanded in powers of $(b - b_0)$:
\beq
	a_{\rm out} = a_{\rm in} + \left[\frac{\log G_0}{2}(b/b_0 - 1) - \frac{G_0 - (1 + \log G_0)}{8} (a_{\rm in}/b_0)^2\right] a_{\rm in} \label{eq:09-nt0}
\eeq
Combining this with (\ref{eq:09-ai-delay2}) and noting that $a_{\rm out} \approx a_{\rm in}$, we get a difference equation for $a_i(t)$.  Below it is written for a single $d$-bit delay:
\bea
	a_i(t+1) - a_i(t) & = & \left[\frac{\log G_0}{2}(b/b_0 - 1) - \frac{G_0 - (1 + \log G_0)}{8} (a_i(t)/b_0)^2\right] a_i(t) \nonumber \\
	& & \qquad +\ r^2 (a_{i-d}(t) - a_i(t)) + \mbox{(noise terms)} \label{eq:09-nt1}
\eea
Near threshold, the field $a_i(t)$ tends to vary slowly in both position and time.  This justifies replacing $a_i(t)$ with a smoothly-varying function $a(x, t)$ and swapping (\ref{eq:09-nt1}) with a PDE.  Ignoring the noise terms, it is:
\bea
	\frac{\partial a}{\partial t} + \frac{1}{2}\frac{\partial a^2}{\partial t^2} & = & \left[\frac{\log G_0}{2}(b/b_0 - 1) - \frac{G_0 - (1 + \log G_0)}{8} \frac{a^2}{b_0^2}\right] a \nonumber \\
	& & \qquad -\ r^2 d \frac{\partial a}{\partial x} + \frac{r^2 d^2}{2} \frac{\partial^2 a}{\partial x^2} \label{eq:09-nt2}
\eea

Steady-state solutions will drift with a speed $v_d = r^2 d$.  Substituting $x = \xi + v_d t$, one obtains a driftless equation of motion which, upon neglecting higher-order time-derivative terms ($\partial_t^2 a, \partial_t\partial_\xi a \ll \partial_t a$), yields:
\beq
	\frac{\partial a}{\partial t} = \left[\frac{\log G_0}{2}(b/b_0 - 1) - \frac{G_0 - (1 + \log G_0)}{8} \frac{a^2}{b_0^2}\right] a + \frac{r^2(1-r^2) d^2}{2} \frac{\partial^2 a}{\partial \xi^2} \label{eq:09-nt3}
\eeq
Although less tractable numerically, the steady state of (\ref{eq:09-nt3}) can be found analytically, yielding helpful insights about domain walls as discussed in the next section.

\section{Collective Dynamics of 1D Chain}
\label{sec:09-coll}

For the fiber OPO Ising machine, the evolution of the 1D chain is a two-stage process: in the {\it growth stage}, the field is weak compared to the saturation value, pump depletion can be ignored and the signal grows exponentially from the vacuum.  Because of inter-pulse coupling, different (Fourier) modes will grow at different rates, the ferromagnetic mode growing fastest.  This lasts for a time $T$, which is logarithmic in the saturation power and inversely proportional to the normalized pump amplitude.

In the {\it saturation stage}, the field saturates to one of two values: $a \rightarrow \pm a_0$.  The sign depends on the sign of the field after the growth stage.  Different regions will have different signs, called {\it domains} in analogy to the classical ferromagnet, and these domains will be separated by topological defects (domain walls).  The domain walls are not fixed, and their mutual attraction can cause some of the smaller domains to annihilate.

\subsection{Growth Stage}
\label{sec:09-growth}

In the growth stage, the field $a_i(t)$ follows Eq.~(\ref{eq:09-linear}).  Restricting attention to the 1D chain using a single delay line, this becomes:
\beq
	a_i(t+1) = G_0^{\frac{1}{2}(b/b_0 - 1)} \left[t^2 a_i + r^2 a_{i-1}\right] + \mbox{(noise terms)} \label{eq:09-linear2}
\eeq
The linear map (\ref{eq:09-linear2}) is diagonalized by going to the Fourier domain $a_i \rightarrow \tilde{a}_k$.  For small $k$, the result is:
\bea
	\tilde{a}_k(t+1) & = & G_0^{\frac{1}{2}(b/b_0-1)}\left(t^2 + r^2 e^{2\pi i k/N}\right) \tilde{a}_k(t) \nonumber \\
	& \approx & \underbrace{G_0^{\frac{1}{2}(b/b_0-1)} e^{-2t^2r^2(\pi k/N)^2}}_{\rm gain}\; \underbrace{\vphantom{G_0^{\frac{1}{2}(b/b_0-1)}} e^{ik(2\pi r^2/N)}}_{\rm drift} \; \tilde{a}_k(t) \label{eq:09-grev}
\eea

\begin{figure}[tbp]
\begin{center}
\includegraphics[width=1.00\textwidth]{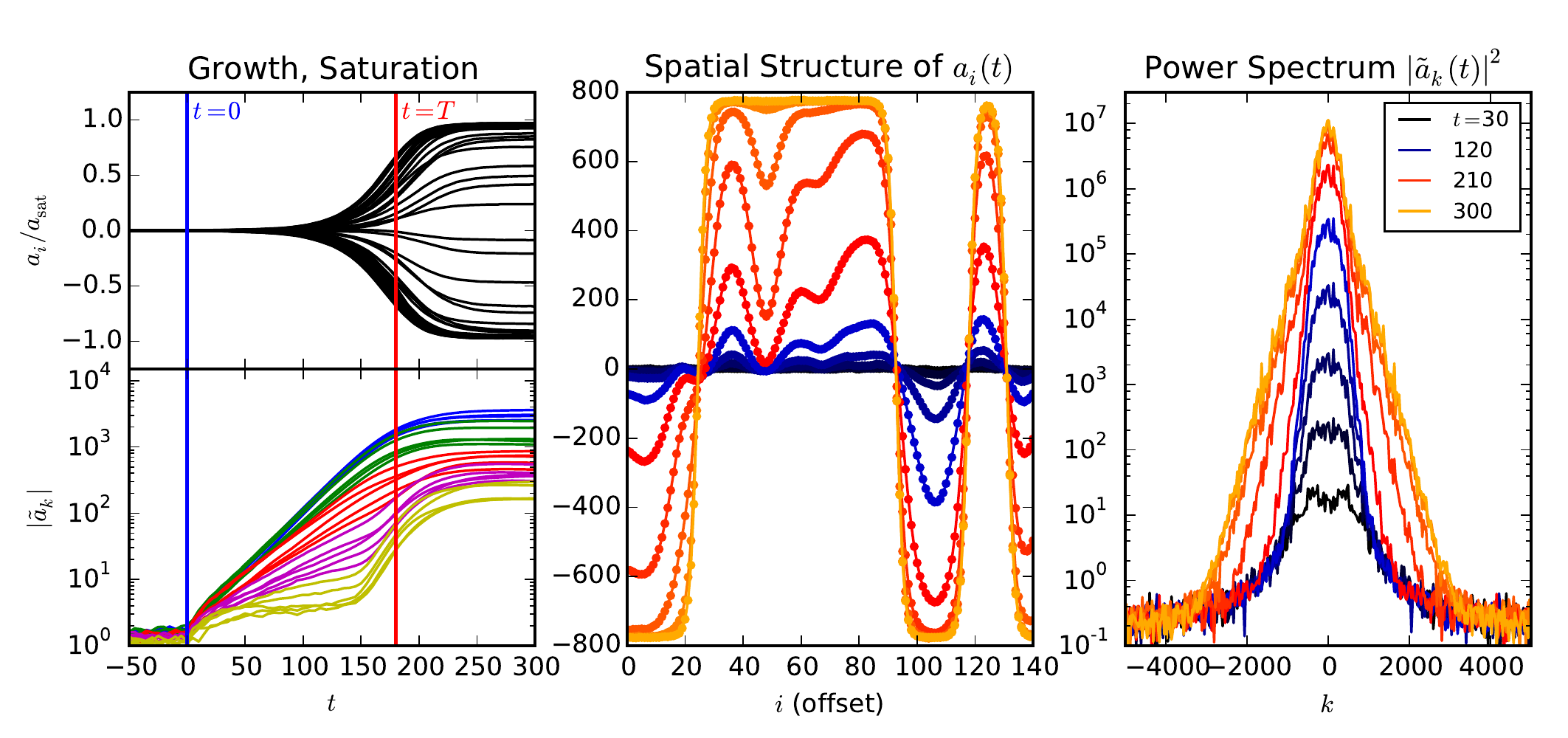}
\caption{Left: growth of OPO pulse amplitudes  $a_i$ (top) and Fourier modes $\tilde{a}_k$ (bottom).  Center: fields $a_i(t)$ for 1D chain at times $t = 30, 60, \ldots 300$ ($x$-axis shifted to cancel drift term).  Right: power spectrum $|\tilde{a}_k|^2$ at times $t = 30, 60, \ldots 300$.  Parameters: $G_0 = 7\;\mbox{dB}$, $b/b_0 = 1.05$}
\label{fig:09-f3}
\end{center}
\end{figure}

The two effects: gain and drift, are separated in Eq.~(\ref{eq:09-grev}).  Drift is a result of the unidirectional coupling.  For a single delay line, the drift speed is $v_d = r^2$.  The gain term depends on $k$, so different modes are amplified at different rates.  This amplification stops when the fields reach their saturation value.  If $N_{\rm sat}$ is the photon number at saturation and we start from vacuum noise, it takes approximately $\log(N_{\rm sat})/\log(G/G_0)$ round trips to reach saturation, that is:
\beq
	T = \frac{1}{b/b_0 - 1} \frac{\log (N_{\rm sat})}{\log(G_0)}
\eeq
Since $T$ depends only logarithmically on $N_{\rm sat}$, which is $O(10^5-10^7)$ in fiber OPOs, factors of two or three are not significant, so we can estimate $N_{sat} \rightarrow b_0^2$, the pump energy at threshold.

Starting with vacuum and propagating the growth equation (\ref{eq:09-grev}) $T$ time steps, we find that at the end of the growth stage the Fourier modes will be distributed as follows:
\beq
	\tilde{a}_k(T) \sim \sqrt{N_{sat}} e^{-2r^2t^2 T(\pi k/N)^2} \label{eq:09-akdist}
\eeq

The modes with smaller $k$ have larger amplitudes, suggesting that the nearest-neighbor interaction forms some kind of short-range order.  A good measure of this is the autocorrelation function $R(x)$.  Before saturation, $R(x)$ is also a Gaussian:
\beq
	R(x) \sim \langle a_i a_{i+x}\rangle = \sum_k e^{2\pi ikx/N} \langle \tilde{a}_k^*\tilde{a}_k \rangle
		\sim e^{-x^2/2x_0^2},\ \ \ x_0 \equiv \sqrt{2T}\,rt \label{eq:09-autocorr}
\eeq

\subsection{Saturation Stage}
\label{sec:09-crst}

In the next stage, pump depletion sets in and the fields inside the OPOs saturate.  The simplest way to model this is to assume that the interaction term $J$ is negligible at this stage.  Under this {\it simple saturation assumption} (SSA), the field in each OPO grows independently until it reaches one of two saturation values: $\pm \sqrt{G/\beta}$.  The {\it sign} of the initial field $c_i(T)$ is preserved, and all its amplitude information is lost.  This can be achieved with a sign function:

\beq
	a_i(\infty) = a_{\rm sat}\,\mbox{sign}[a_i(T)] \label{eq:09-sign}
\eeq

Rather than collapsing into a single ferromagnetic state, the system forms domains of fixed spin, separated by fixed domain walls.  This can be seen in the center plot of Figure \ref{fig:09-f3}.

However, Figure \ref{fig:09-f3} also reveals that the domain walls are not necessarily abrupt phase jumps as (\ref{eq:09-sign}) would have.  Depending on the coupling and pump strength, domain walls can be quite wide.  Near threshold, the shape admits an analytic solution via (\ref{eq:09-nt2}).  Replacing $a_i(t) \rightarrow a(x, t)$ as in Sec.~\ref{sec:09-limits}, a change of variables reduces (\ref{eq:09-nt2}) to the canonical form
\bea
	\frac{\partial\bar{a}}{\partial\bar{t}} & = & (1 - \bar{a}^2)\bar{a} + \frac{1}{2} \frac{\partial^2 \bar{a}}{\partial\bar{x}^2} \label{eq:09-static-norm} \\
	& & \left(\bar{a} = a_0^{-1} a,\ \ \ 
	\bar{x} = (x - vt)/\ell,\ \ \ 
	\bar{t} = t/\tau\right) \label{eq:09-static-norm-units}
\eea
where
\beq
	a_0 = 2b_0 \sqrt{\frac{(b/b_0-1)\log G_0}{G_0 - (1 + \log G_0)}},\ \ \ 
	\ell = \sqrt{\frac{2 r^2(1-r^2)}{(b/b_0-1)\log G_0}},\ \ \ 
	v = r^2,\ \ \ 
	\tau = \frac{2}{(b/b_0-1)\log G_0} \label{eq:09-wallparams}
\eeq
are the saturation field, domain wall length, drift speed, and relaxation time, respectively.  Equation (\ref{eq:09-static-norm}) has an analytic solution: $\bar{c} = \pm \tanh(\bar{x} - \bar{x}_{w})$.  This is the domain wall.  

\begin{figure}[tb]
\begin{center}
\includegraphics[width=1.00\textwidth]{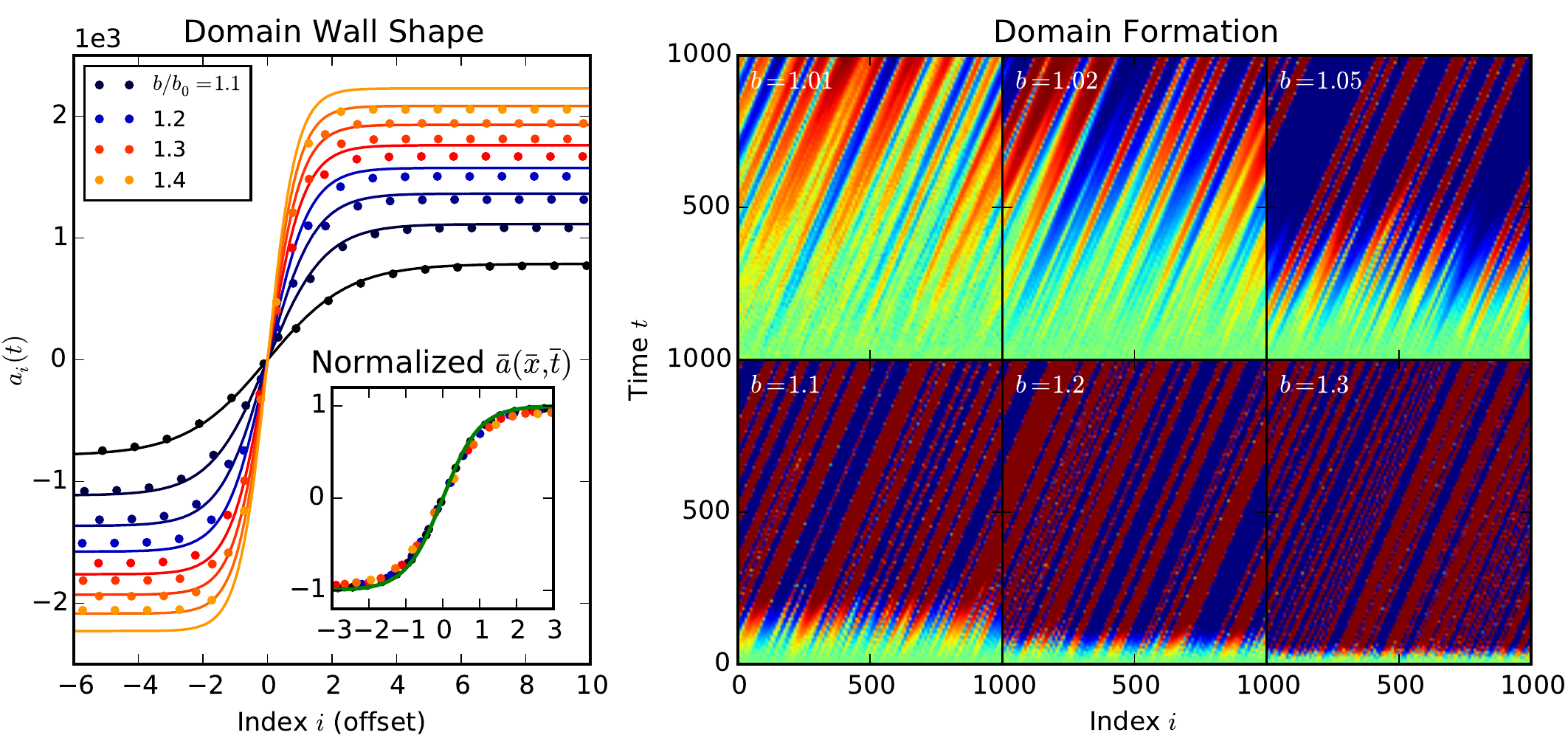}
\caption{Pulse amplitudes $a_i(t)$ near a domain wall as pump is swept slowly from $b/b_0 = 1.0$ to $1.4$ (normalized units in inset).  Right: color plot of pulse amplitudes $a_i(t)$ as function of index $i$ (horizontal) and time $t$ (vertical).  Pump values $b/b_0$ range from $1.01$ to $1.30$.}
\label{fig:09-f5b}
\end{center}
\end{figure}

The left plot of Figure \ref{fig:09-f5b} zooms in on a domain wall.  As the pump grows, the wall gets sharper, its width decreasing as $(b/b_0-1)^{-1/2}$ given in (\ref{eq:09-wallparams}).  If the pump is very strong or the coupling is weak, $\ell \lesssim 1$ and the smoothly-varying field assumption behind (\ref{eq:09-nt2}) breaks down.  However, it seems to hold quite well for the values chosen here (the solid lines in the figure are the $\tanh$ solution).

Domain walls are dynamic objects.  In the presence of a perturbation, they move.  Performing perturbation theory about the $\tanh$ solution, one finds that the Hessian is singular: most of its eigenvalues are $O(1)$ or larger, but for the vector $\partial c/\partial x$, it is zero.  While other perturbations are strongly confined, perturbations along the $\partial c/\partial x$ direction are unimpeded.  These correspond to moving the domain wall left or right.  We can deduce the {\it domain-wall velocity} by taking the inner product (the eigenvalues are orthogonal):
\beq
	\bar{v}_w = -\left[\int{\frac{\partial\bar{a}}{\partial\bar{x}} \frac{\partial\bar{a}}{\partial\bar{x}}d\bar{x}}\right]^{-1} \int{\frac{\partial\bar{a}}{\partial\bar{t}} \frac{\partial\bar{a}}{\partial\bar{x}}d\bar{x}} = -\frac{3}{4} \int{\mbox{sech}^2(\bar{x} - \bar{x}_w) \frac{\partial\bar{a}}{\partial\bar{t}}\,d\bar{x}} \label{eq:09-vw}
\eeq

Consider a function $\bar{a}(x, t)$ with two domain walls at $\pm\bar{L}/2$.  The precise way they are ``glued together'' at $\bar{x} \approx 0$ only matters to second order in the perturbation theory; $a = \tanh(\bar{L}/2 - |\bar{x}|)$ is a valid solution.  Applying (\ref{eq:09-vw}), one finds the following domain-wall speed and collision time:
\beq
	\bar{v}_w = \frac{3}{2}\mbox{sech}^4(\bar{L}/2),\ \ \ 
	\bar{T}_{\bar{L}} = \frac{1}{48}e^{2\bar{L}} \label{eq:09-tcoll}
\eeq

As the domain walls move, smaller domains will evaporate while large domains remain unaffected.  All the domains that survive after a time $t$ have a size $\bar{L} \geq (1/2)\log(48\bar{t})$.

The right plot in Fig.~\ref{fig:09-f5b} shows the formation of domain walls as a color plot in both the pulse index $i$ and time $t$.  The domain drift is obvious here.  In addition, the average domain size clearly shrinks the further the system is from threshold.  Looking closely, one also sees events where domain walls collide and annihilate some of the smaller domains -- but in general this is rare, because the domains that form by time $T$ tend to be moderate in size, and the lifetime (\ref{eq:09-tcoll}) can be quite long.

\section{Final-State Statistics}
\label{sec:statistics}

From the linear- and saturation-stage theory from Section \ref{sec:09-coll}, we can calculate statistical properties of the final-state ($t \rightarrow \infty$) system.  These properties are of interest because can be used to benchmark the performance of different Ising machines, or to compare the Ising machine against other optimizers.  In this section, we compute the autocorrelation function, defect density, success probability and domain-length histogram for the 1D Ising machine.  These are measurable quantities, allowing for a direct comparison between theory and experiment.

\subsection{Autocorrelation Function}

The autocorrelation function, given by $R(x) = \langle a_i a_{i+x} \rangle / \langle a_i^2 \rangle$, is a key quantity in statistical mechanics.  For the thermal Ising model with $H = -\tfrac{1}{2} J \sum_i \sigma_i \sigma_{i+1}$, it falls off exponentially with distance in one dimension, $R(x) = \tanh(\beta J)^x$.

Since the Ising machine is not in thermal equilibrium, we do not expect {\it a priori} that $R(x)$ will be exponential.  Indeed, at the end of the growth stage, Eq.~(\ref{eq:09-autocorr}) shows that $R(x)$ is a Gaussian.  The easiest way to compute $R(x)$ as $t \rightarrow \infty$ is to assume the simple saturation approximation (\ref{eq:09-sign}).  Replacing $a_i \rightarrow \mbox{sign}(a_i)$, the autocorrelation at $t \rightarrow \infty$ is found to be:
\beq
	R(x;\infty) = 1 - 2P(a_{i}(T)a_{i+x}(T) < 0) \label{eq:09-autocorr2}
\eeq
where $T$ is the saturation time.  Since the evolution in $t < T$ is approximately linear, the probability distribution of $a(T)$ is a two-dimensional Gaussian.  Its covariance is related to the autocorrelation at time $T$, $R(x;T) = e^{-x^2/2x_0^2}$:
\beq
	\sigma_{i,i+x} = \begin{bmatrix} 1 & e^{-x^2/2x_0^2} \\ e^{-x^2/2x_0^2} & 1 \end{bmatrix}
\eeq
Following (\ref{eq:09-autocorr2}), the autocorrelation may be expressed as an integral over a Gaussian with linear constraints:
\beq
	R(x) = 1 - 4 \int_{\mathcal{Q}} \frac{1}{2\pi\sqrt{\det\sigma}} e^{-\frac{1}{2}a^T\sigma^{-1}a} da_i da_{i+x} \label{eq:09-autocorr3}
\eeq
where $\mathcal{Q} = \{(a_i, a_{i+x}): a_i < 0, a_{i+x} > 0\}$ is the upper-left quadrant in $(a_i, a_{i+x})$.  To solve this, perform a linear transformation that diagonalizes the quadratic form in the exponent; $\mathcal{Q}$ is deformed to a pie slice, and the resulting integral is proportional to its angle.  The autocorrelation becomes:
\beq
	R(x) = 1 - \frac{4}{\pi} \tan^{-1}\sqrt{\tanh(x^2/4x_0^2)} \label{eq:09-rx-final}
\eeq

\begin{figure}[tbp]
\begin{center}
\includegraphics[width=1.0\textwidth]{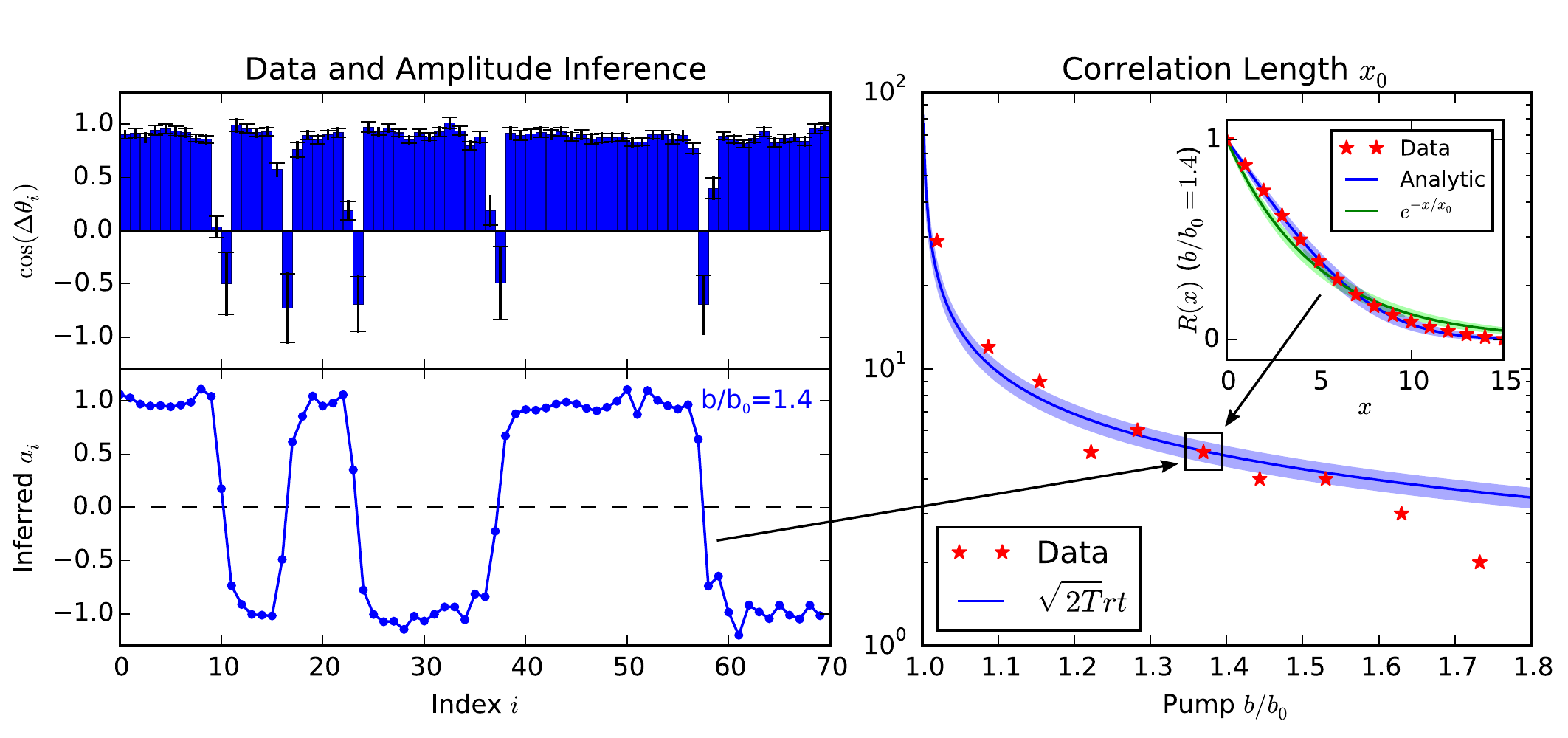}
\caption{Left: fiber OPO data for $\cos(\Delta\theta_i)$ (top) and reconstructed amplitude $a_i$ (bottom).  Right: autocorrelation length as a function of pump $b/b_0$, compared to Eq.~(\ref{eq:09-autocorr}).  Inset: autocorrelation $R(x)$ and analytic fits: form (\ref{eq:09-rx-final}) in blue, exponential in green.  Red stars are experimental data.  Shaded regions show sensitivity of the analytic curves to $N_{\rm sat}$ when varied from $4 \times 10^5$ to $4 \times 10^7$.}
\label{fig:09-f6}
\end{center}
\end{figure}

To compute $R(x)$ from the experimental data, one must first reconstruct the pulse amplitudes $a_i(t)$ from the measurement record.  In Inagaki et al.\cite{Inagaki2016}, no local oscillator is present, so the signal is passed through a Mach-Zehnder with a delay line, measuring the quantities $I_{1,i} = |a_i + a_{i+1}|^2$, $I_{2,i} = |a_i + a_{i+1}|^2$.  If the pulse energy $|a_i|^2$ is the same for each pulse, the angle between neighboring pulses is given by $\cos(\Delta\theta_i) = (I_{1,i}-I_{2,i})/(I_{1,i}+I_{2,i})$.  A negative value of $\cos(\Delta\theta_i)$ indicates a phase flip.  This is plotted in the upper-left panel of Fig.~\ref{fig:09-f6}.  Taking $a_i$ to be real for the degenerate OPO, we can invert the relation between the $a_i$ and the $I_{1,i}, I_{2,i}$ to reconstruct the original amplitude sequence $a_i(t)$.  It is then straightforward to compute the autocorrelation function and the correlation length.

The right plot of Fig.~\ref{fig:09-f6} shows the autocorrelation length as a function of pump amplitude, obtained by fitting experimental data to (\ref{eq:09-rx-final}).  The experimental $x_0$ agree with Eq.~(\ref{eq:09-autocorr}), with a particular fit for $b/b_0 \approx 1.4$ shown in the inset.

Although, Eq.~(\ref{eq:09-rx-final}) looks like an exponential to the unaided eye, plotting them on top of each other, the former is a much better fit to the experimental data, as shown in the inset plot.  However, it turns out that the best exponential fit to (\ref{eq:09-rx-final}) is $R(x) = e^{-x/x'_0}$, with $x'_0 = 1.00463x_0$.  Thus, we can obtain $x_0$ from experimental data by fitting the autocorrelation to an exponential.  The right plot in \ref{fig:09-f6} shows this for a variety of pump powers.  The agreement with experimental data is reasonably good.

\subsection{Defect Density}

Another key statistic is the {\it defect (domain wall) density}.  This is the average number of domain walls divided by the size of the chain $n_d = N_d/N$.  The average domain length is then $L_d = 1/n_d$.  For a thermal Ising model with $H = -\tfrac{1}{2} J \sum_i \sigma_i\sigma_{i+1}$, one has $n_d = (1+e^{\beta J})^{-1}$.

Since $a_i(\infty)$ has fixed amplitude, one can compute $n_d$ from the autocorrelation function: $n_d = (1 - R(1))/2$.  For $x_0 \gtrsim 10$, $R(x)$ may be linearized about $x = 0$, giving the result:

\beq
	n_d = \frac{1}{\pi x_0},\ \ \ L_d = \pi x_0,\ \ \ x_0 = \sqrt{2T}\,rt \label{eq:09-nda}
\eeq

\begin{figure}[tbp]
\begin{center}
\includegraphics[width=1.00\textwidth]{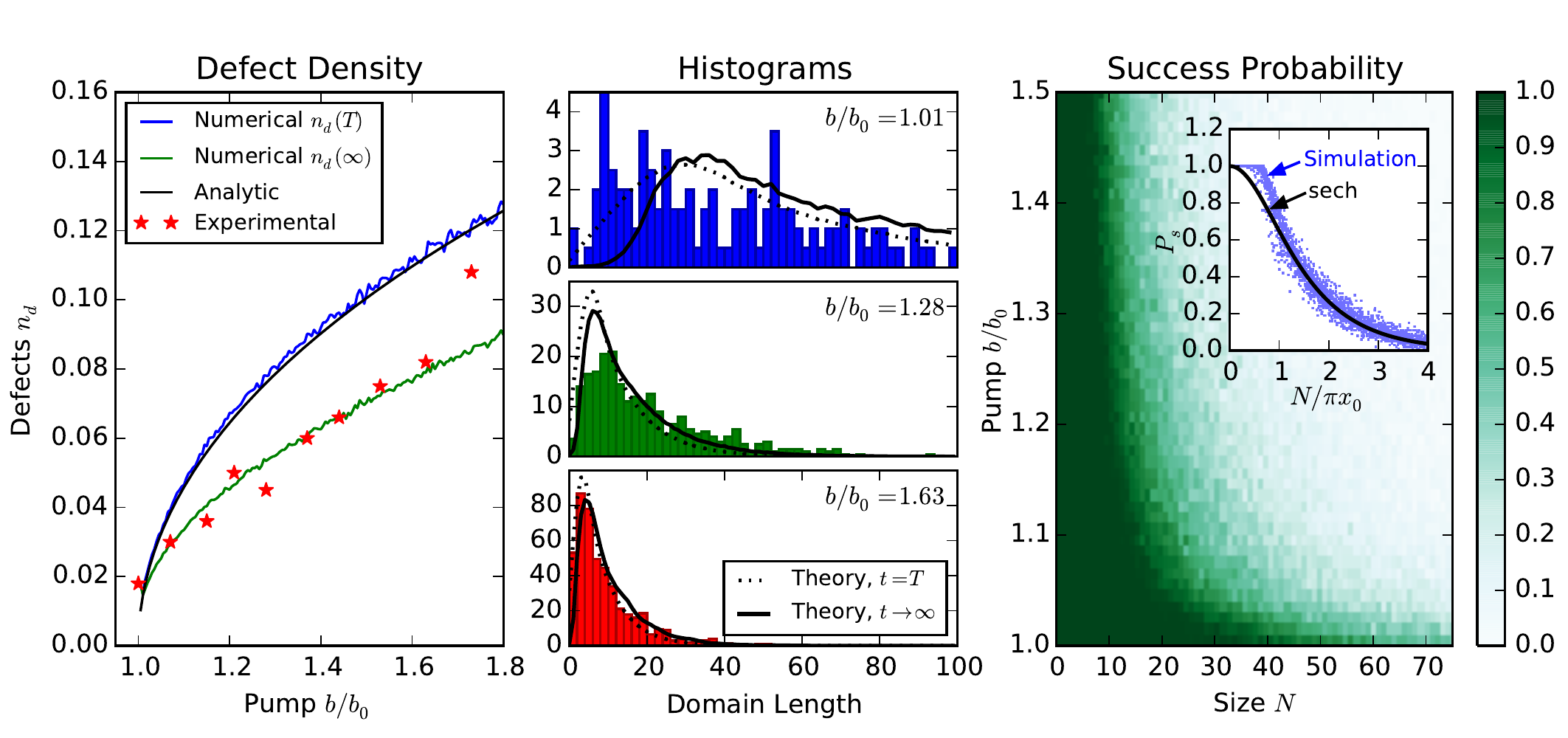}
\caption{Left: plot of defect density as a function of pump $b/b_0$, numerical and analytic models (Eq.~\ref{eq:09-nda}) compared to experimental data.  Center: domain length histograms for $b/b_0 = 1.01$, $1.28$ and $1.63$.  Bars denote experimental data.  Right: success probability $P_s$ as a function of system size $N$ and pump $b/b_0$.}
\label{fig:09-f7}
\end{center}
\end{figure}

Figure \ref{fig:09-f7} (left) compares experimental data from Inagaki et al.\cite{Inagaki2016} (Fig. 3) to both Eq.~(\ref{eq:09-nda}) and numerical simulations.  The data match the simulations when $t \rightarrow \infty$, but deviate from Eq.~(\ref{eq:09-nda}).  This suggests that the full numerical model works well, but Eq.~(\ref{eq:09-nda}), which relies on the simple saturation assumption (\ref{eq:09-sign}), is inaccurate.  This is the result of domain-wall motion and collision in the saturation stage, which reduces the number of defects as $t \rightarrow \infty$.

\subsection{Domain Length Histograms}

Experimental data for the domain-length distribution $P(\ell)$ is plotted in Fig.~\ref{fig:09-f7} (center).  There is a reasonable fit between the data and numerical simulations as $t \rightarrow \infty$.  Note, however, that the calculated histogram at $t \rightarrow \infty$ differs from that at $t \rightarrow T$.  This difference reflects the domain-wall dynamics in the saturation phase.  In particular, since small domains evaporate faster than large domains, the population of small domains is depleted, and the average domain length grows.  Since $N_d L_d = N$, an increase in domain length results in a decrease in defect density, giving rise to the difference between the $t = T$ and $t = \infty$ lines in the left plot.

In a thermal Ising model, the probability distribution of spin $\sigma_{i+1}$ depends only on its nearest neighbors; mathematically this makes it a Markov chain in $i$.  Thus, the distribution $P(\ell)$ should be exponential in $L$: $P(\ell) \sim e^{-n_d \ell}$.  The histograms in Fig.~\ref{fig:09-f7} have exponential tails, but are clearly not exponential for $\ell$ near zero.  This means that the Ising machine never reaches thermal equilibrium, even when $t \rightarrow \infty$.  Rather, it ``freezes out'' fluctuations accumulated during the linear growth stage, through a highly nonlinear process involving domain wall motion and collisions.  Only if one waits an exponentially long time will the larger domains evaporate, bringing the machine to the ground state.

\subsection{Success Probability}

\begin{table}[bt]
\tbl{Comparison of thermal Ising model and the final state in the coherent Ising machine.}
{\begin{tabular}{p{0.2\textwidth}|p{0.25\textwidth}|p{0.25\textwidth}|p{0.2\textwidth}}
\Hline
  & Thermal 
  & CIM Theory 
  & Experiment \\
\hline
Mechanism 
  & Thermally-activated flips create a Boltzmann distribution.
  & Linear growth of OPO amplitudes, followed by saturation.
  & \\ \hline
Correlation $R(x)$
  & $e^{-x/x_0}$
  & See Eq.~(\ref{eq:09-rx-final})
  & Matches CIM \\
Corr.\ length $x_0$
  & $-1/\log(\tanh(\beta J/2))$
  & $\sqrt{2T} rt$
  & Matches CIM \\
Defect density $n_d$
  & $1/(1+e^{\beta J})$ 
  & $1/\pi x_0$
  & Matches CIM \\
Length dist.\ $P(\ell)$
  & $(1+e^{-\beta J})^{-\ell}$
  & Non-exponential, Fig.~\ref{fig:09-f7}
  & Matches CIM \\
Success probability
  & $\mbox{sech}(N e^{-\beta J})$
  & See Fig.~\ref{fig:09-f7}
  & $\approx 0$ for $N = 10000$ \\
\Hline
\end{tabular}}
\label{tab:09-t2}
\end{table}

The success probability $P_s$ of the Ising machine is defined as the probability that it reaches the ground state at some time $T_{\rm final}$.  The chosen $T_{\rm final}$ depends on experimental parameters, should be large compared to the saturation time $T$, but not exponentially large (since this would always give the ground state).  Figure \ref{fig:09-f7} (right) plots the success probability (numerically computed for $t \rightarrow \infty$) as a function of system size and pump power.  As expected, the probability is greatest near threshold for small systems, where the average defect number $N/\pi x_0$ is small.

If we assume the final state is thermal, the success probability can be calculated analytically.  For an $N$-spin ring with $H = -\tfrac{1}{2} J \sum_i \sigma_i\sigma_{i+1} + N/2$ and periodic boundary conditions, the partition function is:
\beq
	Z = \sum_n (1 + (-1)^n) e^{-n \beta J} = (1 + e^{-\beta J})^N + (1 - e^{-\beta J})^N
\eeq
The success probability is the ground-state probability for the system.  The ground state has energy zero and degeneracy 2, so $P_s = 2/Z$.  For low defect densities, $e^{-\beta J} \ll 1$, and the success probability becomes:
\beq
	P_s = \mbox{sech}\left(N e^{-\beta J}\right) \label{eq:ps-sech}
\eeq
Note that $e^{-\beta J}$ is the approximate defect density (for $e^{-\beta J} \ll 1$).  Thus, in analogy to the thermal model, we suspect that the success probability of the 1D Ising machine should depend on the defect density as well.  Using the relation $n_d \approx 1/\pi x_0$, in the inset figure, $P_s$ is plotted against $N/\pi x_0$ for all $(N, b/b_0)$ values shown in the larger plot.  Like the thermal model, the full Ising machine success probability falls off exponentially for high $N$, fitting reasonably well to the form $P_s = \mbox{sech}(N/\pi x_0)$.

\section{2D and Frustrated Systems}
\label{sec:09-2d}

\subsection{2D Square Lattice}
\label{sec:09-2dlat}

The two-dimensional Ising lattice exhibits richer physics than its 1D counterpart.  In particular, the thermal 2D system has a phase transition at finite temperature with long-range order below the transition temperature\cite{Onsager1944}.  Likewise, we suspect that a mechanism must exist to ensure long-range order in the 2D Ising machine.

Referring back to Figure \ref{fig:09-f2}, an $m \times n$ Ising lattice can be realized in an OPO network using a 1-bit and $m$-bit delay.  This implements the couplings $a_{i,i} \rightarrow a_{i,i+1}$, $a_{i,i}\rightarrow a_{i+1,i}$.  If the spins are serialized in C order $a_{i,j} \leftrightarrow a_{mi+j}$, then periodic boundary conditions $a_{i,n+1} = a_{i,1}$; $a_{m+1,j} = a_{a,j+1}$ are enforced.  There is a slight vertical offset compared to standard periodic boundary conditions (see Fig.~\ref{fig:09-f2}) but in the limit $m, n \gg 1$ with ferromagnetic couplings, this offset is negligible.

As before, the dynamics are described by a {\it growth stage} and a {\it saturation stage}.  In the growth stage, the Fourier modes are amplified independently, in analogy to Eq.~(\ref{eq:09-akdist}) we have:
\beq
	\tilde{a}_k(T) \sim \sqrt{N_{sat}} e^{-2r^2t^2 T \pi^2 (k_x^2+k_y^2)/N^2} \label{eq:09-akdist2}
\eeq

This gives the same autocorrelation function, generalized to two dimensions: $R(x) = e^{-(x^2+y^2)/2x_0^2}$, with $x_0 = \sqrt{2T}\,rt$.  Here $T = (b/b_0-1)^{-1} \log(N_{\rm sat})/\log(G_0)$ is the saturation time; see Sec.~\ref{sec:09-growth}.

\begin{figure}[tbp]
\begin{center}
\includegraphics[width=1.0\textwidth]{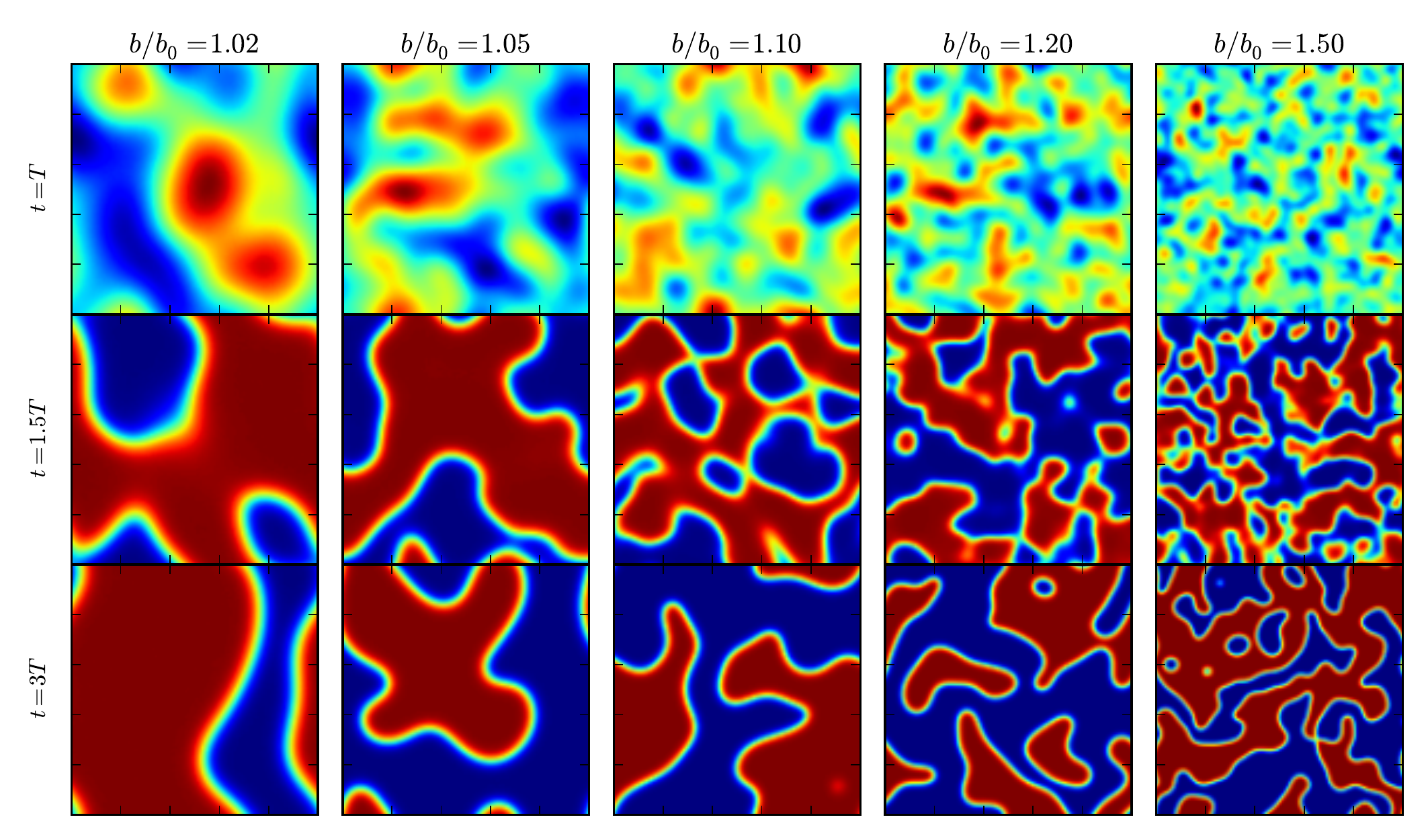}
\caption{Simulation of 2D OPO Ising machine, $100\times 100$ grid.  Pump ranges from $b/b_0 = 1.02$ to $1.50$.}
\label{fig:09-f9}
\end{center}
\end{figure}

Growth-stage fluctuations are imprinted on the domain structure of the OPO, and persist for some time.  Since these fluctuations are longer-range the larger the saturation time $T$, the Ising machine displays longer-range order when the pump is closer to threshold, just like the 1D case.  Figure \ref{fig:09-f9} shows the state of the machine for five different pump powers $b/b_0$.  The larger $b/b_0$, the smaller the domains that form.

After saturation, we can proceed analytically as long as the pump is near threshold.  Invoking the limit (\ref{eq:09-nt0}) and inserting both horizontal and vertical delays to obtain the two-dimensional analog of (\ref{eq:09-nt2}):
\bea
	a_{i,j}(t+1) - a_{i,j}(t) & = & \left[\frac{\log G_0}{2}(b/b_0 - 1) - \frac{G_0 - (1 + \log G_0)}{8} (a_{i,j}(t)/b_0)^2\right] a_{i,j}(t) \nonumber \\
	& & + \left[r^4 a_{i-1,j-1}(t) + r^2t^2 (a_{i-1,j}(t)+a_{i,j-1}(t)) - t^4 a_{i,j}(t)\right] \label{eq:09-nt4}
\eea

Near threshold, the field $a_{ij}(t)$ tends to vary slowly in both position and time.  Following the same procedures used to obtain (\ref{eq:09-nt3}), replaces the discrete increments with derivatives and drops higher-order $\partial^2a/\partial t^2, \partial^2a/\partial x\partial t, \partial^2a/\partial y\partial t$ terms, obtaining:
\beq
	\frac{\partial a}{\partial t} = \left[\frac{\log G_0}{2}(b/b_0 - 1) - \frac{G_0 - (1 + \log G_0)}{8} \frac{a^2}{b_0^2}\right] a + \frac{r^2(1-r^2)}{2} \left[\frac{\partial^2 a}{\partial \xi_x^2} + \frac{\partial^2 a}{\partial \xi_y^2}\right] \label{eq:09-nt5}
\eeq
where $\xi_x = x - r^2 t$, $\xi_y = y - r^2 t$ are the comoving coordinates.  Setting $\bar{a} = a_0^{-1} a$, $\bar{x} = (x-vt)/\ell$, $\bar{y} = (y-vt)/\ell$, $\bar{t} = t/\tau$, Eq.~(\ref{eq:09-nt5}) is converted to its canonical form (the 2D version of (\ref{eq:09-static-norm}))
\beq
	\frac{\partial\bar{a}}{\partial\bar{t}} = (1 - \bar{a}^2)\bar{a} + \frac{1}{2} \left(\frac{\partial^2 \bar{a}}{\partial\bar{x}^2} + \frac{\partial^2 \bar{a}}{\partial\bar{y}^2}\right) \label{eq:09-static-norm2}
\eeq
with $a_0$, $\ell$, $v$ and $\tau$ given in (\ref{eq:09-wallparams}).  The steady-state solutions to (\ref{eq:09-static-norm2}), $\bar{a} = \tanh(\bar{x}\cos\theta + \bar{y}\sin\theta)$, are linear domain walls.

Curved domain walls will move towards the center of curvature at a rate proportional to $1/R$.  This can be seen intuitively if we imagine each spin on the wall picking a sign based on a majority vote of its neighbors.  The rate can be computed by considering the special case of a circular domain.  Working in cylindrical coordinates, (\ref{eq:09-static-norm2}) becomes:
\beq
	\frac{\partial\bar{a}}{\partial\bar{t}} = (1 - \bar{a}^2)\bar{a} + \frac{1}{2} \frac{\partial^2 \bar{a}}{\partial\bar{r}^2} + \frac{1}{2\bar{r}} \frac{\partial \bar{a}}{\partial\bar{r}} \label{eq:09-static-norm3}
\eeq

Here, $(2r)^{-1} \partial\bar{a}/\partial\bar{r}$ is a perturbation to the 1D equation (\ref{eq:09-static-norm}).  Applying the same results used to compute the attraction of neighboring walls (Eq.~(\ref{eq:09-vw})), the drift velocity is
\beq
	\bar{v}_w = -\frac{1}{2\bar{r}} \label{eq:09-vw-2d}
\eeq

\begin{figure}[btp]
\begin{center}
\includegraphics[width=1.00\textwidth]{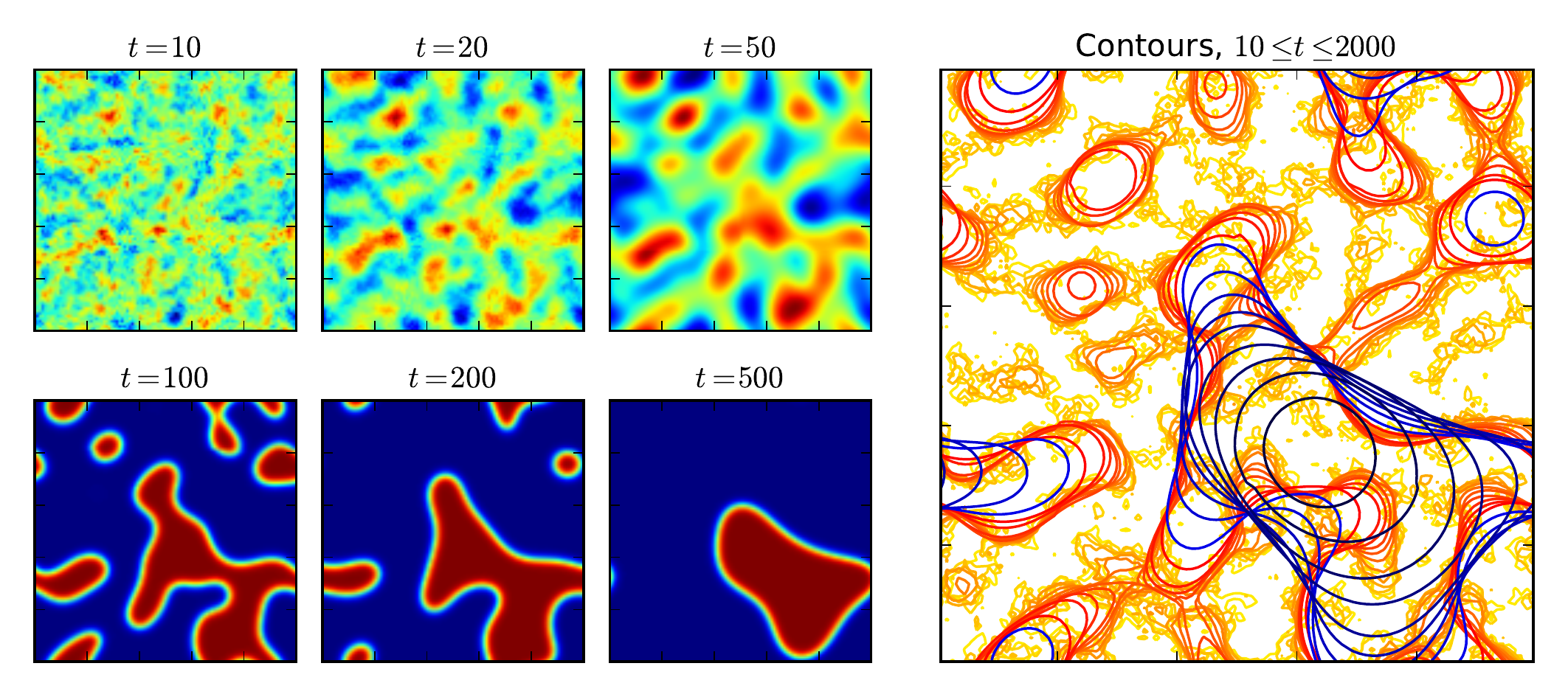}
\caption{Left: Simulation of 2D OPO Ising machine, $100\times 100$ grid, $b/b_0 = 1.1$.  Right: Location of domain walls for $10 \leq t \leq 2000$ (larger $t$ values are darker).}
\label{fig:09-f8}
\end{center}
\end{figure}

For a circular domain of size $\bar{r}$, this gives $d\bar{r}/dt = -1/2\bar{r}$, which implies a collapse time of $\bar{T}_{\bar{r}} = \bar{r}^2$.  This time only scales quadratically with the domain size -- unlike the 1D case, where domains of size $L$ live for $(1/48)e^{2\bar{L}}$ time.  As a result, the 2D Ising machine on an $m\times n$ lattice should reach the ground state with high probability if allowed to run for $O(m^2, n^2)$ time.

Figure \ref{fig:09-f8} shows a simulation for $b/b_0 = 1.1$ ($T \approx 100$).  The top-left plots correspond to linear growth, which by $t = 100$ has saturated into domains.  Locally, the domain walls migrate towards their center of curvature, which the more tightly curved parts moving faster, following (\ref{eq:09-vw-2d}).  This can also be seen in the right plot, which superimposes the domain boundaries 23 time slices in $10 \leq t \leq 2000$.  Just after $t = 2000$, the system collapses into the ferromagnetic state.

\subsection{Frustrated Chains and Lattices}

In frustrated Ising models, different couplings compete and the resulting spin structure can be much richer than simple (anti-)ferromagnetism.  Most systems in classical and quantum physics involve frustration to some degree.  Moreover, frustration is a intimately connected to computational complexity; while non-frustrated Ising problems are trivial to solve, frustration makes the problem NP-hard in general\cite{Barahona1982}.

The simplest way to introduce frustration to the 1D Ising chain is to cascade two 1-bit delays, one with phase 0 (beamsplitter $r = \sqrt{J_1}$) and one with phase $\pi$ ($r = \sqrt{J_2}$).  (This requires tunable beamsplitters, but the tuning only needs to happen on slow timescales.)  During the linear growth stage, the round-trip gain is:
\beq
    a(t+1) = G_0^{\frac{1}{2}(b/b_0-1)} \left[a_i(t) + (J_1 - J_2) a_{i-1}(t) - J_1 J_2 a_{i-2}(t)\right] \label{eq:09-fr-dadt}
\eeq

The nearest-neighbor coupling is ferromagnetic if $J_1 > J_2$, antiferromagnetic if $J_2 > J_1$, but in either case it wants to align next-nearest neighbors.  This conflicts with the next-nearest term in (\ref{eq:09-fr-dadt}), causing frustration.  (The case $J_1 = J_2$ is special because the nearest-neighbor term cancels out.  In this case, the even and odd spins decouple, so the chain can be ``unwrapped'' into two independent (antiferromagnetic) chains of size $N/2$).

Because of the time-invariant couplings, the eigenvectors will be Fourier modes.  For Fourier mode $k$, we have:
\beq
	G_k \equiv \left| \frac{\tilde{a}_k(t+1)}{\tilde{a}_k(t)} \right|^2 = G_0^{b/b_0-1} 
    \left[1 - 4J_1(1-J_1)\sin^2(k/2N)\right]\left[1 - 4J_2(1-J_2)\cos^2(k/2N)\right]
\eeq
There are three distinct possibilities:

\begin{figure}[tbp]
\begin{center}
\includegraphics[width=1.0\textwidth]{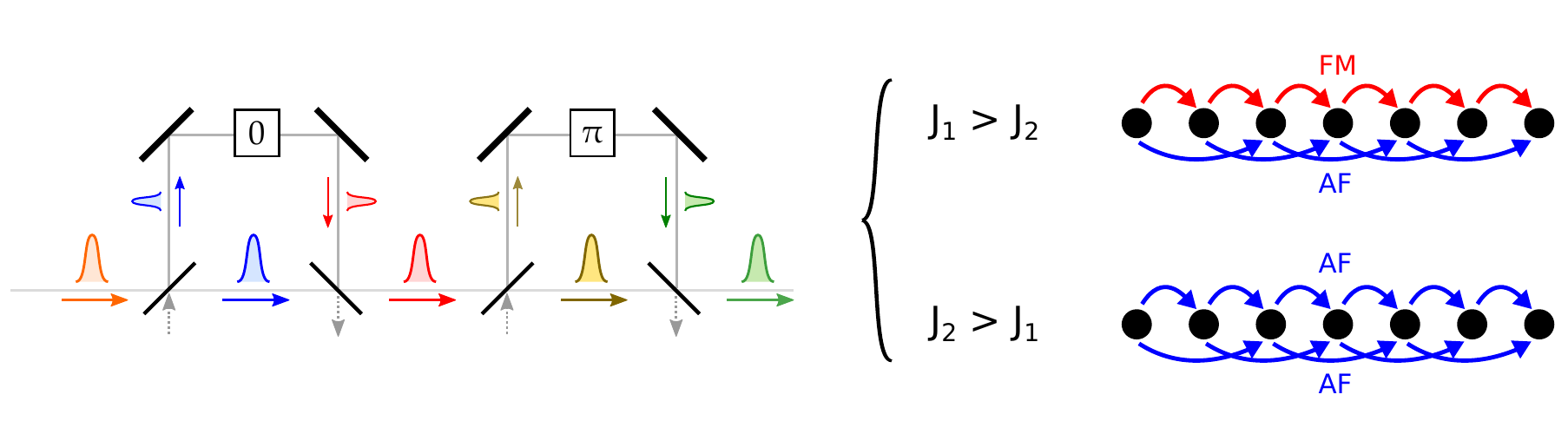}
\caption{Cascade of ferro- and antiferromagnetic couplings creates a frustrated spin chain.}
\label{fig:09-f11}
\end{center}
\end{figure}

\begin{enumerate}
    \item $G_k$ decreasing for all $k \in [0, \pi]$.  Maximum at $k = 0$.  Ferromagnetic order at growth stage.
    \item $G_k$ increasing on $k \in [0, \pi]$.  Maximum at $k = N\pi$.  Antiferromagnetic order at growth stage.
    \item $G_k$ non-monotonic.  Maximum for some $k \in (0, N\pi)$.  Frustrated system at growth stage.
\end{enumerate}
Examining the first derivatives of $G_k$ at $x = \{0, 1\}$, we deduce that:
\begin{eqnarray}
    J_2(1-J_2) < \frac{J_1(1-J_1)}{1 + 4J_1(1-J_1)} & \Leftrightarrow & \mbox{Ferromagnetic} \nonumber \\
    J_1(1-J_1) < \frac{J_2(1-J_2)}{1 + 4J_2(1-J_2)} & \Leftrightarrow & \mbox{Antiferromagnetic} \label{eq:09-phasediag}
\end{eqnarray}
Figure \ref{fig:09-f11} (left plot) illustrates the phase diagram defined by (\ref{eq:09-phasediag}).  

For weak couplings, there is a clear boundary between ferro- and antiferromagnetic behavior.  But for strong couplings, we get this interesting ``frustrated'' regime.  It's not hard to show that, in the frustrated regime, the $k$ with maximum gain is:
\beq
    k_{\rm max} = \cos^{-1}\left(\frac{J_1(1-J_1) - J_2(1-J_2)}{4J_1(1-J_1)J_2(1-J_2)}\right)
\eeq
Note that this is only defined in the frustrated region; elsewhere $k_{\rm max}$ is $0$ or $\pi$ depending on whether the dominant coupling is ferro- or antiferromagnetic.

Frustration increases the threshold beyond $b_0$, the uncoupled OPO threshold.  The gain at pump $b$ for the dominant mode is calculated to be:

\beq
	G_{\rm max} = \left\{\begin{array}{ll}
		G_0^{b/b_0-1} \left[1 - 4J_2(1 - J_2)\right] & \mbox{FM} \\
		G_0^{b/b_0-1} \left[1 - 4J_1(1 - J_1)\right] & \mbox{AF} \\
		G_0^{b/b_0-1} \frac{\left(J_1(1-J_1) + J_2(1-J_2) - 4J_1(1-J_1)J_2(1-J_2)\right)^2}{4J_1(1-J_1)J_2(1-J_2)} & \mbox{Frustrated} \end{array}\right.
\eeq

The gain still varies exponentially with $b$.  One can define the frustrated threshold $b'_0$ such that $G_{\rm max}(b'_0) = 1$, and threshold gain $G'_0 = G_{\rm max}(0)^{-1}$.  In terms of these quantities, the gain at $k_{\rm max}$ varies as $G_{\rm max} = (G'_0)^{b/b'_0-1}$, in analogy to (\ref{eq:09-gain})
		
\begin{figure}[tbp]
\begin{center}
\includegraphics[width=1.00\textwidth]{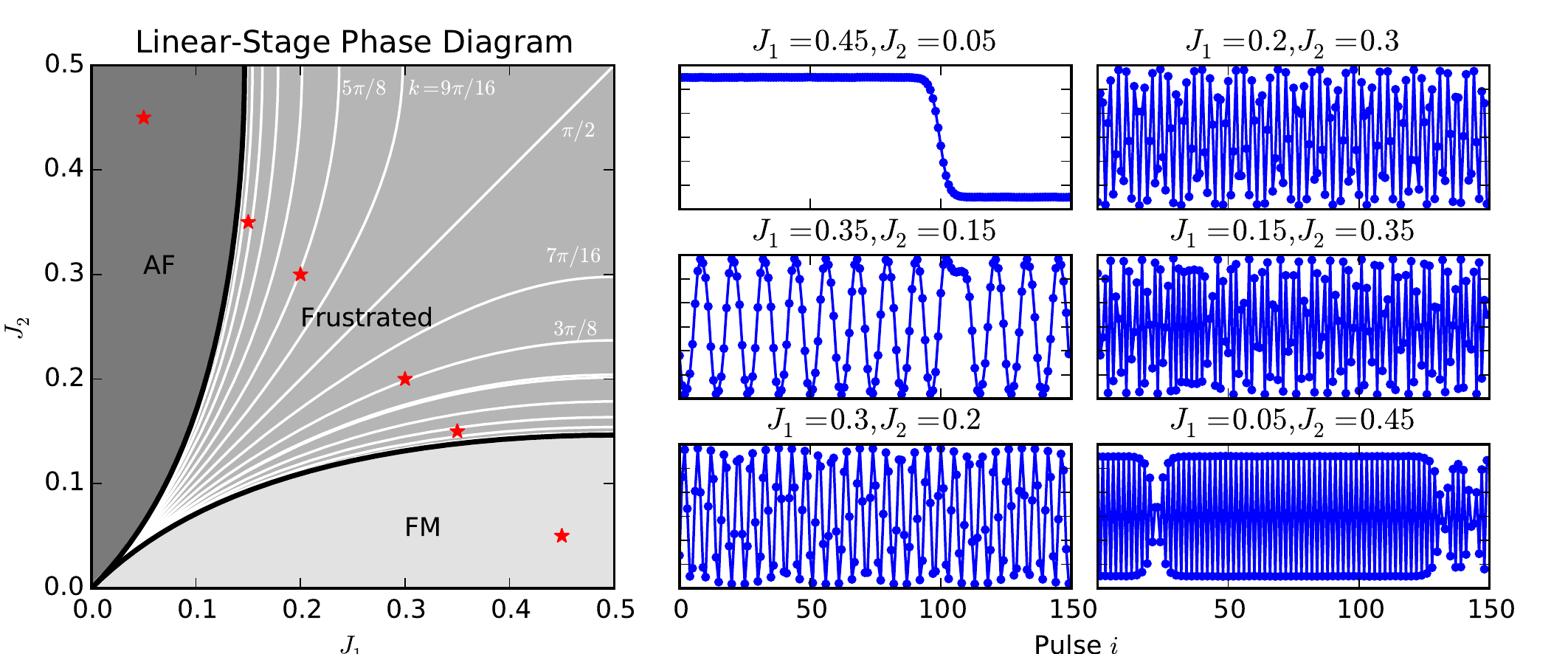}
\caption{Left: phase diagram for the frustrated chain.  Contours of $k_{\rm max}$ shown in white.  Red stars correspond to plots on the right.  Right: Ising machine output as the couplings $J_1$ and $J_2$ are varied.}
\label{fig:09-f11}
\end{center}
\end{figure}

The right plots in Figure \ref{fig:09-f11} show the transition from ferromagnetic to antiferromagnetic order as one passes through the frustration region in parameter space.  First, the ferromagnetic domains give way to an oscillatory order parameter, whose wavelength decreases until it starts to approximate antiferromagnetic order.  Eventually this leads to the antiferromagnetic domains in the lower-right plot.  The $J_1 > J_2$ and $J_2 > J_1$ regimes are related by a symmetry: replacing $J_1 \leftrightarrow J_2$ and $a_i \rightarrow (-1)^i a_i$, the equations of motion are unchanged.

Most of the theory developed above carries over to frustrated 2D lattices.  The phase diagram in Fig.~\ref{fig:09-f11} is unchanged, but now the Fourier modes in the growth stage have two wavenumbers $k_x, k_y$.  It is not hard to show that the mode gain is:
\begin{eqnarray}
G_k & = & G_0^{b/b_0-1} \left[1 - 4J_1(1-J_1)\sin^2(k_x/2N)\right]
    \left[1 - 4J_2(1-J_2)\cos^2(k_x/2N)\right] \nonumber \\
    & & \qquad \times\left[1 - 4J_1(1-J_1)\sin^2(k_y/2N)\right]
    \left[1 - 4J_2(1-J_2)\cos^2(k_y/2N)\right]
\end{eqnarray}

\begin{figure}[tbp]
\begin{center}
\includegraphics[width=1.00\textwidth]{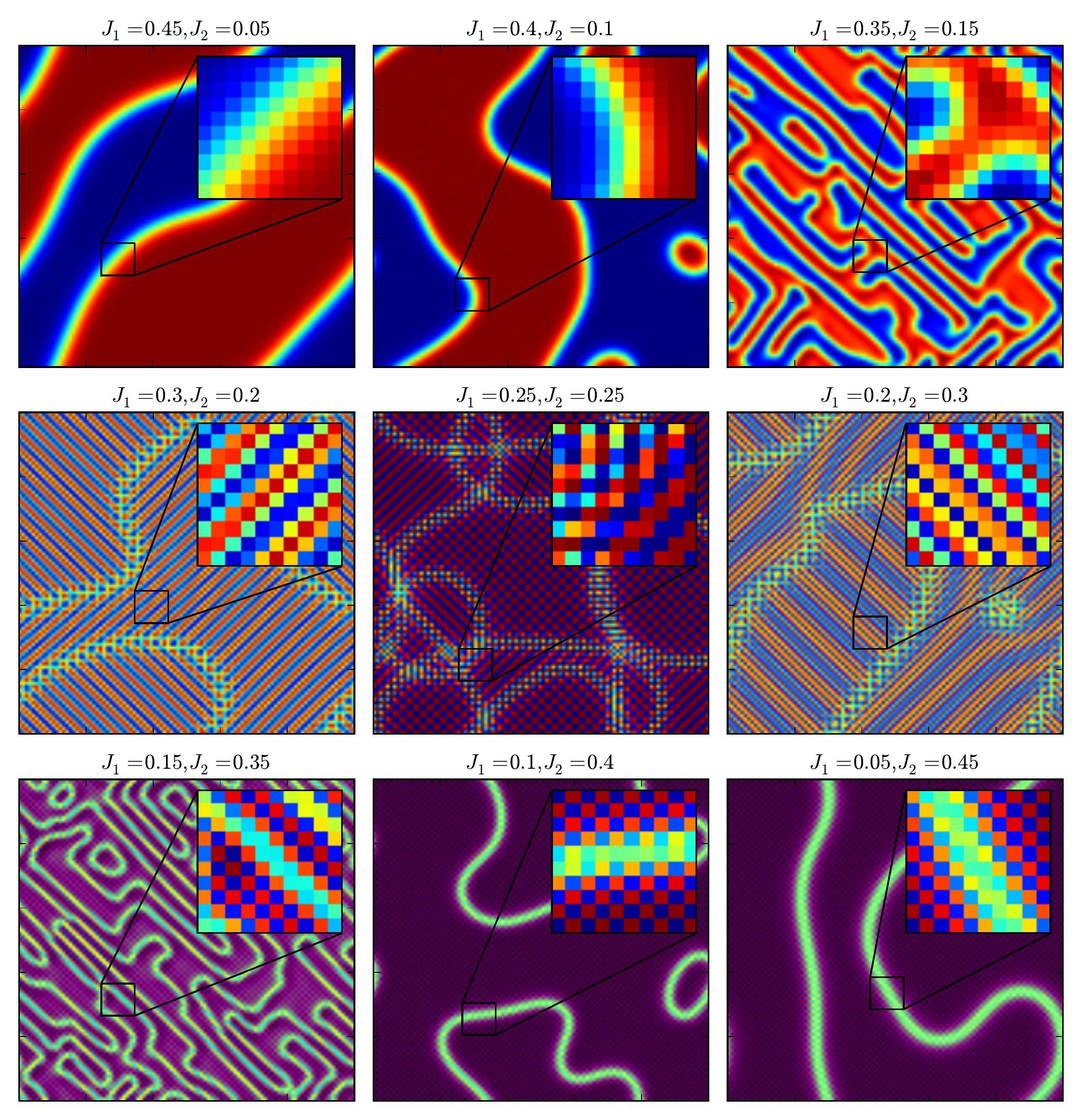}
\caption{Ising machine simulations ($b/b'_0 = 1.02$, $t=1500$) for frustrated system with $J_1, J_2$ ranging from mostly ferromagnetic (upper left) to mostly antiferromagnetic (bottom right)}
\label{fig:09-f12}
\end{center}
\end{figure}

There are four frequencies that maximize $G_k$ are $(k_{\rm max}, k_{\rm max})$, $(-k_{\rm max}, -k_{\rm max})$, $(k_{\rm max}, -k_{\rm max})$, $(-k_{\rm max}, k_{\rm max})$.  These create upper and lower diagonal ``stripes'' (see Fig.~\ref{fig:09-f12}).  These striped domains compete with each other and form domains with domain walls in the frustrated region.

In addition to the stripes, Figure \ref{fig:09-f12} shows some interesting behavior in and near the frustrated zone.  At $J_1 = J_2$, one finds a doubling of the unit cell and three domain types appear to exist: stripes with $k=(\pm\pi/2, \pm\pi/2)$ and checkerboards. One should not read too much into this, because there is no nearest-neighbor coupling in this system, meaning it is equivalent to four $50\times 50$ lattices with antiferromagnetic coupling, interleaved in both $x$ and $y$ directions.

Near the frustration transition, the lattice forms filaments of opposite phase.  The ferromagnetic domain walls are subject to two nearly-equal opposing forces: nearest-neighbor interactions want to shrink and circularize the walls, as per Fig.~\ref{fig:09-f8}; on the contrary, the next-nearest neighbor effect wants to create striped order in the system.  The resulting structure is not unlike that of the manganites, where opposite phases coexist and percolate into each other\cite{Dagotto2013,Nagaev2002}.

\section{XY Machine Based on OPO}
\label{sec:xy}

\begin{figure}[b]
\begin{center}
\includegraphics[width=0.90\textwidth]{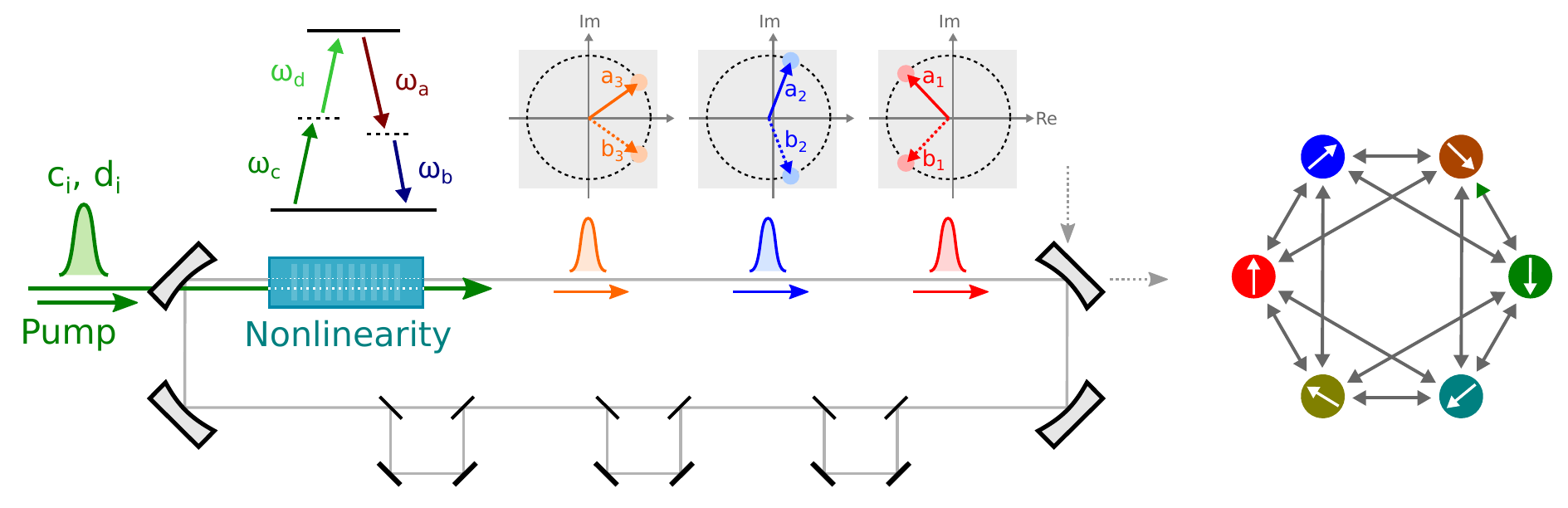}
\caption{Diagram of a time-multiplexed XY machine based on nondegenerate four-wave mixing.}
\label{fig:09-f13}
\end{center}
\end{figure}

In statistical physics, the XY model refers to a network of spins $\vec{\sigma}_i \in \mathcal{R}^2$, $|\vec{\sigma}_i^2| = 1$, with the Hamiltonian $U(\sigma) = -\tfrac{1}{2} \sum_{ij} J_{ij}\vec{\sigma}_i \cdot \vec{\sigma}_j$.  Each spin has a continuous $U(1)$ degree of freedom, rather than being discrete-valued.  It could equivalently be formulated in terms of angles, with $\vec{\sigma}_i = (\cos\phi_i, \sin\phi_i)$ living in a potential
\beq
	U(\phi) = -\frac{1}{2} \sum_{ij}{J_{ij} \cos(\phi_i - \phi_j)} \label{eq:09-uxy}
\eeq
Nondegenerate OPOs also have a $U(1)$ degree of freedom above threshold.  As a result, it is natural to map XY dynamics onto a nondegenerate OPO.  

\subsection{Gain Equations}

Consider a non-degenerate fiber OPO based on the four-wave mixing process $\omega_a + \omega_b \leftrightarrow \omega_c + \omega_d$ (Fig.~\ref{fig:09-f13}).

\begin{align}
	\frac{\d a}{\d z} = \frac{1}{2}\gamma b^* c d - \frac{1}{2}\alpha a & & 
		\frac{\d c}{\d z} = -\frac{1}{2}\gamma ab d^* - \frac{1}{2}\alpha c \\		
	\frac{\d b}{\d z} = \frac{1}{2}\gamma a^* cd - \frac{1}{2}\alpha b & & 
		\frac{\d d}{\d z} = -\frac{1}{2}\gamma ab c^* - \frac{1}{2}\alpha d
\end{align}

The pump fields $c$ and $d$ do not resonate.  We can assume without loss of generality that they are real.  The signal experiences gain when the phases of $a$ and $b$ are equal and opposite, that is $ab \in \mathbb{R}$.  If $(a, b)$ is a steady state in the OPO, so is $(a\,e^{i\phi}, b\,e^{-i\phi})$.  Thus the nondegenerate OPO has a ring of steady states, each with its own phase.  The spin $\vec{\sigma}_i$ is represented with this phase.

Rescaling $a, b, c, d, z$ to eliminate $\gamma$ and $\alpha$, the field equations are reduced to their canonical form.  Assuming $ab \in \mathbb{R}$ and real $c, d$:

\beq
	\frac{\d |\bar{a}|}{\d s} = |\bar{b}| \bar{c}\bar{d},\ \ \ 
	\frac{\d |\bar{b}|}{\d s} = |\bar{a}| \bar{c}\bar{d},\ \ \ 
	\frac{\d \bar{c}}{\d s} = -|\bar{a}||\bar{b}| \bar{d},\ \ \ 
	\frac{\d \bar{d}}{\d s} = -|\bar{a}||\bar{b}| \bar{c} \label{eq:09-xy-eom1}
\eeq

To proceed further, one assumes that one of the pump fields is much stronger than the other: $|d| \gg |c|$.  This allows us to ignore depletion in $d$ and treat $\bar{d}$ as a constant.  It is worth noting that the resulting system (in the $\alpha = 0$ limit) becomes equivalent to a $\chi^{(2)}$ OPO, with $\tfrac{1}{2}\gamma d \rightarrow \epsilon$ the $\chi^{(2)}$ coupling parameter.

The general three- and four-wave mixing problems can be solved analytically in terms of Jacobi elliptic functions\cite{Armstrong1962,Chen1989-2,Chen1989}.  Generally, two limits are of interest for OPOs: singly- and doubly-resonant.

\subsubsection{Singly Resonant Case}

For the singly resonant OPO, the initial idler amplitude is zero.  Following Armstrong et al.\cite{Armstrong1962} (Eq.~6.13) and including fiber and additional linear losses, the output signal is:
\beq
	a_{\rm out} = a_{\rm in} G_0^{-1/2} \sqrt{1 + \frac{c_{\rm in}^2}{a_{\rm in}^2}\left[1 - \mbox{cd}^2\left(\frac{1}{2}\gamma d_{\rm in} L_{\rm eff} \sqrt{a_{\rm in}^2 + c_{\rm in}^2}; \frac{c_{\rm in}^2}{a_{\rm in}^2 + c_{\rm in}^2}\right)\right]}
\eeq
where $\mbox{cd}(x;v)$ is a Jacobi elliptic function.  This linearizes for $a \ll c, d$ to $a_{\rm out} = G_0^{-1/2} \cosh(\tfrac{1}{2}\gamma c_{\rm in} d_{\rm in} L_{\rm eff}) a_{\rm in}$.  Since cavity losses are included here, the threshold $c = c_0$ is defined so that $a_{\rm out} = a_{\rm in}$.  The gain above threshold is:
\beq
	G = \cosh\left(\frac{c}{c_0} \cosh^{-1}(G_0^{1/2})\right)^2
\eeq

\subsubsection{Doubly Resonant Case}

If the signal and idler frequencies are similar enough and we don't filter one of them out, they will propagate through the cavity with the same $Q$ factor.  As a result, $a$ and $b$ will have the same magnitude.  If, furthermore, the overall phase is stabilized, then we have $a = b^*$.  All modes orthogonal to the $a = b^*$ subspace experience loss in the gain medium, and can be ignored.

Setting $b_{\rm in} = a_{\rm in}^*$ amounts to equating the constants of motion $A, B$.  Equations (\ref{eq:09-xy-eom1}) reduce to $\d \bar{a}/\d s = \bar{d}_{\rm in}\,\bar{a}\sqrt{\bar{c}_{\rm in}^2 + \bar{a}_{\rm in}^2 - \bar{a}^2}$.  This matches Eq.~(\ref{eq:09-dads}) up to scaling factors, so the input-output relation is analogous.  Converting to the form (\ref{eq:09-aout}), we have:
\beq
	a_{\rm out} = a_{\rm in} G_0^{\frac{1}{2}\sqrt{a_{\rm in}^2+c_{\rm in}^2}/c_0} \left[1 + (G_0^{\sqrt{a_{\rm in}^2+c_{\rm in}^2}/c_0} - 1) \frac{1 - \sqrt{1 - (a_{\rm in}/\sqrt{a_{\rm in}^2+c_{\rm in}^2})^2}}{2}\right]^{-1} \label{eq:09-inout-xy}
\eeq

Linearizing (\ref{eq:09-inout-xy}) for small input fields, we find $a_{\rm out} = G_0^{c/2c_0} a_{\rm in}$.  Thus the (power) gain for the waveguide pumped above threshold is the same as in the degenerate case:
\beq
	G = G_0^{c/c_0}
\eeq
Going to third order in $a_{\rm in}$, it is not hard to derive the XY version of Eq.~(\ref{eq:09-inout3}), valid when the pump is near threshold:
\beq
	a_{\rm out} = a_{\rm in} \sqrt{G/G_0} \left[1 - \frac{G - (1 + \log G)}{4} \frac{|a_{\rm in}|^2}{c^2} + O\left((a_{\rm in}/b)^4\right)\right] \label{eq:09-inout3-xy}
\eeq

As in the degenerate case, quantum noise can be modeled by adding vacuum fluctuations to the input pump fields $c_{i,\rm in} \rightarrow c_{i,\rm in} + w_i^{(c)}$, $d_{\rm in} \rightarrow d_{i,\rm in} + w_i^{(d)}$ (and idler $b_{\rm in} \rightarrow b_{i,\rm in} + w_i^{(b)}$, if the system is singly resonant) and treating the loss in signal as a lumped element after the gain medium: $a_{i,\rm out} \rightarrow a_{i,\rm out} + \sqrt{1 - 1/G_0}w_i^{(a)}$ (plus $b_{i,\rm out} \rightarrow b_{i,\rm out} + \sqrt{1 - 1/G_0}w_i^{(b)}$ if doubly resonant).  As before, $w$ is a discrete-time noise process with vacuum statistics: $\langle w^* w \rangle = \tfrac{1}{2}$.

In the rest of this paper, we assume a doubly-resonant OPO for concreteness.  Because the signal and idler amplitudes are equal, the results are analytically more tractable.  But it is worth noting that the same calculations could be done using the singly-resonant results above.

\subsection{Couplings}

Tracing the paths in the canonical delay-line diagram (Fig.~\ref{fig:09-f2}), vacuum enters the cavity through the first beamsplitter.  The transmitted beam passes along the cavity without delay, while the reflected beam is delayed by one pulse spacing, contributing to $a_{i+d}$ instead.  There are five parameters: $r, t, r', t', \phi$, which can in principle vary in time.

\beq
	a_i \rightarrow t_i t'_i a_i + r_i r'_{i-d} e^{i\phi_i} a_{i-d} + \left(t_i r_i w^{(J)}_i + r_i t_{i-d} e^{i\phi}w^{(J)}_{i-d}\right) \label{eq:09-ai-delay-xy}
\eeq

Couplings will be more difficult to implement in the doubly-resonant regime because both signal and idler fields propagate with separate parameters $r, t, r', t', \phi$.  To maintain the condition $a = b^*$ the beamsplitter coefficients must be the same and the phases must be opposite:

\bea
	a_i & \rightarrow & t_i t'_i a_i + r_i r'_{i-d} e^{i\phi_i} a_{i-d} + \left(t_i r_i w^{(J,a)}_i + r_i t_{i-d} e^{i\phi}w^{(J,a)}_{i-d}\right) \label{eq:09-ai-delay-xy1} \nonumber \\
	b_i & \rightarrow & t_i t'_i b_i + r_i r'_{i-d} e^{-i\phi_i} b_{i-d} + \left(t_i r_i w^{(J,b)}_i + r_i t_{i-d} e^{-i\phi}w^{(J,b)}_{i-d}\right) \label{eq:09-ai-delay-xy2}
\eea

\section{1D and 2D XY Models}
\label{sec:xy1d2d}

Like the Ising machine, the XY machine is a dynamical system whose motion can be divided into two stages.  In the {\it growth stage}, quantum fluctuations are amplified from the vacuum.  The modes that are amplified the most are the largest eigenvalues of the coupling matrix $J_{ij}$.  The growth stage runs for a time $T = O((c-c_0)^{-1})$; the longer $T$, the more the state is resolved to the largest eigenvectors.  In the {\it Kuramoto stage}, the spin amplitude saturates and the system follows Kuramoto-model dynamics, which may be highly nonlinear.  After a while, it relaxes to a local minimum of the potential (\ref{eq:09-uxy}).

Both 1D and 2D XY models are intimately connected to the topology of $U(1)$.  Since the first homotopy group of the $d$-dimensional torus is $\pi_1(T_d) = \mathbb{Z}^d$, local minima are given by {\it winding states} which can be characterized by $d$ winding numbers $w_1, w_2, \ldots w_d$ for the $d$ dimensions\cite{NakaharaBook}.  In addition, in $d \geq 2$ dimensions, topologically protected vortices can form, which in thermal systems give rise to the Berezinskii-Kosterlitz-Thouless (BKT) vortex-pair transition\cite{Kosterlitz1973}.

\subsection{1D Chain}
\label{sec:xy1d}

A ferromagnetic 1D chain is realized with a single delay line of phase 0; see Sec.~\ref{sec:09-coll}.  The linear dynamics of $a_i(t)$ are the same as for the Ising model: working in the Fourier basis $\tilde{a}_k(t)$, the system of difference equations diagonalizes.  The initial quantum noise is amplified to macroscopic values.  At the saturation time $T = (c/c_0-1)^{-1} \log(N_{\rm sat})/\log(G_0)$, these Fourier modes have mean amplitude:
\beq
	\sqrt{\langle |\tilde{a}_k(T)|^2 \rangle} = \sqrt{N_{\rm sat}} e^{-2r^2 t^2 T(\pi k/N)^2} \label{eq:09-ak-xy}
\eeq

This has a correlation length $x_0 = \sqrt{2T}\,rt$.  The only difference here is both quadratures of $a$ experience gain in the XY model.  After the growth stage, the amplitude $a_i$ quickly saturates, but the phase is still free to move.  Assuming $a_i(t) = a_{\rm sat} e^{i\phi_i(t)}$, the phase is found to follow the difference equation:
\beq
	\phi_i(t+1) = \left[t^2 \phi_i + r^2 \phi_{i-1}\right] \label{eq:09-km-xy}
\eeq
Equation (\ref{eq:09-km-xy}) is a linear equation with the boundary condition $\phi_{N} = \phi_0 + 2m\pi$.  As in the growth stage, the best way to solve it is to use a Fourier series:
\beq
	\phi_x = \frac{m x}{N} + \sum_k \phi_k e^{2\pi ikx/N}
\eeq
Note that (\ref{eq:09-km-xy}) and (\ref{eq:09-linear2}) are the same up to the constant gain term.  Thus the eigenvalues for the $\phi_k$ will be:
\beq
	\lambda_k = \frac{\phi_i(t+1)}{\phi_i(t)} = t^2 + r^2 e^{-2\pi ik/N} = \underbrace{e^{-\frac{1}{2}(r t)^2(2\pi k/N)^2}}_{\rm diffusion}\, \underbrace{e^{ir^2(2\pi k/N)}}_{\rm drift} 
\eeq

\begin{figure}[tbp]
\begin{center}
\includegraphics[width=1.00\textwidth]{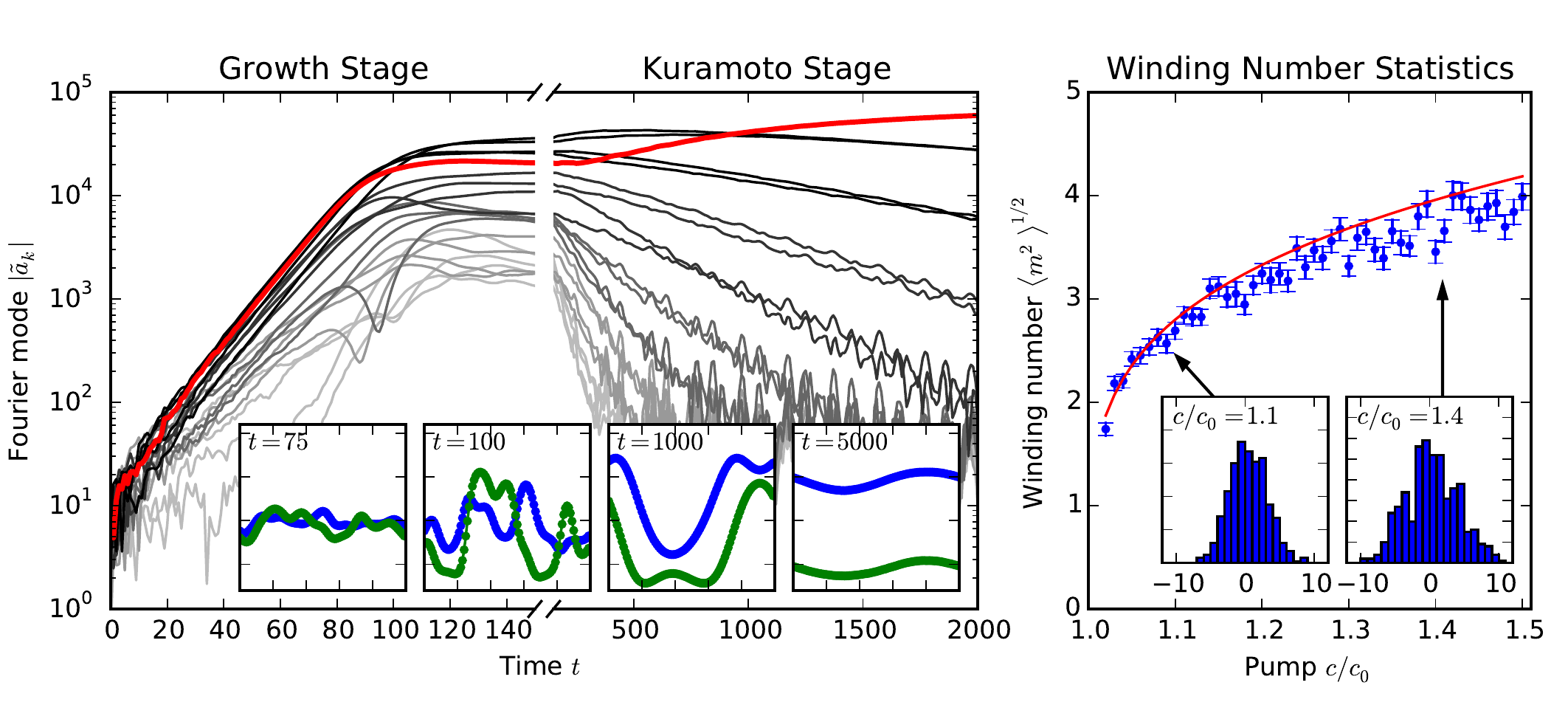}
\caption{Left: Fourier modes $|\tilde{a}_k(t)|$ for a 100-spin XY machine simulation.  Right: average winding number $\langle m^2\rangle^{1/2}$, with the fit $0.232\sqrt{N/x_0}$.}
\label{default}
\end{center}
\end{figure}

The steady state will be a state of constant winding $\phi_x = mx/N$.  For $m = 0$ this is the global minimum for the XY potential, for $m \neq 0$ an excited local minimum.  For sufficiently long chains, where $N \gg x_0$, the parts of the chain separated by $\gtrsim x_0$ are uncorrelated, so on these length-scales, the phase executes a random walk.  Thus the average number of windings is normally distributed about zero, with a standard deviation that goes as $\langle m^2\rangle^{1/2} \approx 0.232 \sqrt{N/x_0}$ (the constant must be determined numerically).

Note the two timescales in this problem.  The first is the growth-stage time.  If we want to reach the global minimum, the growth stage must be long enough for $x_0 \approx N$.  Since $x_0 = \sqrt{2T} rt$, this constrains the growth-stage time to be $T \gtrsim (N/rt)^2$. On the other hand, just to reach a local minimum, we must wait long enough in the Kuramoto stage for the phase excitations $\phi_k$ to decay to zero -- this takes $O(N/rt)^2$ time as well.  So no matter what kind of minimum we want, global or local, we must wait $O(N/rt)$ time, but to get the global minimum, this must happen in the growth stage, when the field is weak compared to saturation.

Another important thing to note is that the precise form of the nonlinear input-output map $a_{\rm in} \rightarrow a_{\rm out}$ does not matter.  In the growth stage, this map is linearized so all that matters is the gain, which determines the saturation time.  In the Kuramoto stage, since the amplitude saturates much more rapidly than the phase dynamics, the equation for $\phi_i$ does not even depend on the gain element.  This seems to suggest that all XY machines are equivalent when it comes to solving the 1D Ising problem.

\subsection{2D Lattice}

As far as local minima are concerned, the 2D lattice is just like a 1D chain in two directions.  The spins are indexed by two coordinates $a_{i,j}$ with the connections $a_{i,i}\rightarrow a_{i,i+1}$, $a_{i,i}\rightarrow a_{i+1,i}$ (Sec.~\ref{sec:09-2dlat}) and the equilibrium solutions are states of constant winding number: $a_{x,y} = e^{i(m_x x + m_y y)}$.  The growth stage is also analogous:  the Fourier amplitudes grow according to (\ref{eq:09-ak-xy}) so that the autocorrelation is $e^{-(x^2+y^2)/2x_0^2}$.

Having saturated the amplitude and thus reached the Kuramoto stage, the 2D model becomes quite different.  Topological {\it vortex defects} form and the dynamics are dominated by inter-vortex interactions.

\subsubsection{Vortex Shape, Frequency}

For an infinite lattice, an isolated vortex is a stable solution to the round-trip equations of the OPO.  Following the analysis leading to (\ref{eq:09-nt3}) and (\ref{eq:09-nt5}), which is applicable in the near-threshold limit, the round-trip equations can be rewritten as a nonlinear PDE with gain and diffusion:
\beq
	\frac{\partial a}{\partial t} = \left[\frac{\log G_0}{2}(c/c_0 - 1) - \frac{G_0 - (1 + \log G_0)}{4} \frac{|a|^2}{c_0^2}\right] a + \frac{r^2(1-r^2)}{2} \left[\frac{\partial^2 a}{\partial \xi_x^2} + \frac{\partial^2 a}{\partial \xi_y^2}\right] \label{eq:09-nt6}
\eeq
where $\xi_x = x-r^2 t$, $\xi_y = y-r^2t$.  This differs from (\ref{eq:09-nt5}) only in that $a$ is complex-valued here.  Setting $\bar{a} = a_0^{-1} a$, $\bar{x} = (x-vt)/\ell$, $\bar{y} = (y-vt)/\ell$, $\bar{t} = t/\tau$, this equation is converted to its canonical form:
\beq
	\frac{\partial\bar{a}}{\partial\bar{t}} = (1 - |\bar{a}|^2)\bar{a} + \frac{1}{2} \left(\frac{\partial^2 \bar{a}}{\partial\bar{x}^2} + \frac{\partial^2 \bar{a}}{\partial\bar{y}^2}\right) \label{eq:09-static-norm2}
\eeq
with the constants
\begin{align}
	a_0 &= c_0 \sqrt{2\frac{(c/c_0-1)\log G_0}{G_0 - (1 + \log G_0)}},
	& \ell &= \sqrt{\frac{2 r^2(1-r^2)}{(c/c_0-1)\log G_0}}, \nonumber \\
	v &= r^2,
	& \tau &= \frac{2}{(c/c_0-1)\log G_0} \label{eq:09-wallparams2}
\end{align}
Going to polar coordinates $(r, \phi)$, the vortex is the solution $A(r) e^{\pm i\phi}$ with $A(r)$ satisfying the differential equation:
\beq
	\frac{1}{2} \left(A'' + \frac{1}{r} A'\right) + \left(1 - A^2 - \frac{1}{2r^2}\right)A = 0 \label{eq:09-vorta}
\eeq

It turns out that $A(r) \approx \tanh (r)$ is a good approximation for the amplitude.  This comes from the fact that $\tanh (r)$ is a solution to (\ref{eq:09-vorta}) if the $A'/2r$ and $A/2r^2$ terms are ignored, and these terms nearly cancel out for the solution $\tanh(r)$.

To calculate the number of vortices at time $T$, one finds the probability that the phase winds $2\pi$ around one unit cell of the lattice.  Defining $a_{\rm sq} = [a_P, a_Q, a_R, a_S]$ as the pulse amplitudes for four corners of any lattice cell, clockwise ordered, the joint probability is a Gaussian with the covariance matrix:
\beq
	\avg{a_{\rm sq}a_{\rm sq}^T} = \begin{bmatrix} 1 & e^{-1/2x_0^2} & e^{-1/x_0^2} & e^{-1/2x_0^2} \\
		e^{-1/2x_0^2} & 1 & e^{-1/2x_0^2} & e^{-1/x_0^2} \\
		e^{-1/x_0^2} & e^{-1/2x_0^2} & 1 & e^{-1/2x_0^2} \\
		e^{-1/2x_0^2} & e^{-1/x_0^2} & e^{-1/2x_0^2} & 1 \end{bmatrix}
\eeq
In the near-threshold limit, $x_0 \gg 1$.  Conditioned on the mean value $\mu_{\rm sq} = \tfrac{1}{4}(a_P+a_Q+a_R+a_S)$, $\hat{a}_{\rm sq} \equiv \sqrt{2}\,x_0 a$ is distributed as:
\beq
	P(\hat{a}_{\rm sq} | \mu) = N\left(\sqrt{2}\,x_0\mu \begin{bmatrix} 1 \\ 1 \\ 1 \\ 1 \end{bmatrix},\ 
		\begin{bmatrix} 1 & 0 & -1 & 0 \\ 0 & 1 & 0 & -1 \\ -1 & 0 & 1 & 0 \\ 0 & -1 & 0 & 1 \end{bmatrix}\right)
\eeq
up to terms small in the expansion in $1/x_0$.  The probability $P({\rm vortex})$ depends only on $\xi \equiv |\sqrt{2}\,x_0\mu|$ and is maximal for $\xi = 0$, decaying to zero as $\xi \rightarrow \pm\infty$.  The vortex density is thus 
\beq
	n_v \equiv P({\rm vortex}) = \int{P({\rm vortex}|\sqrt{2}\,x_0\mu) P(\mu) \d^2\mu} \approx \frac{1}{2x_0^2} \int{P({\rm vortex}|\xi) \xi\,\d\xi} \approx \frac{0.159}{x_0^2} \label{eq:09-nvort}
\eeq
The constant $0.159$ in (\ref{eq:09-nvort}) must be determined numerically.  The total number of vortices will be $N n_v$, where $N$ is the size of the lattice.

\begin{figure}[tbp]
\begin{center}
\includegraphics[width=1.0\textwidth]{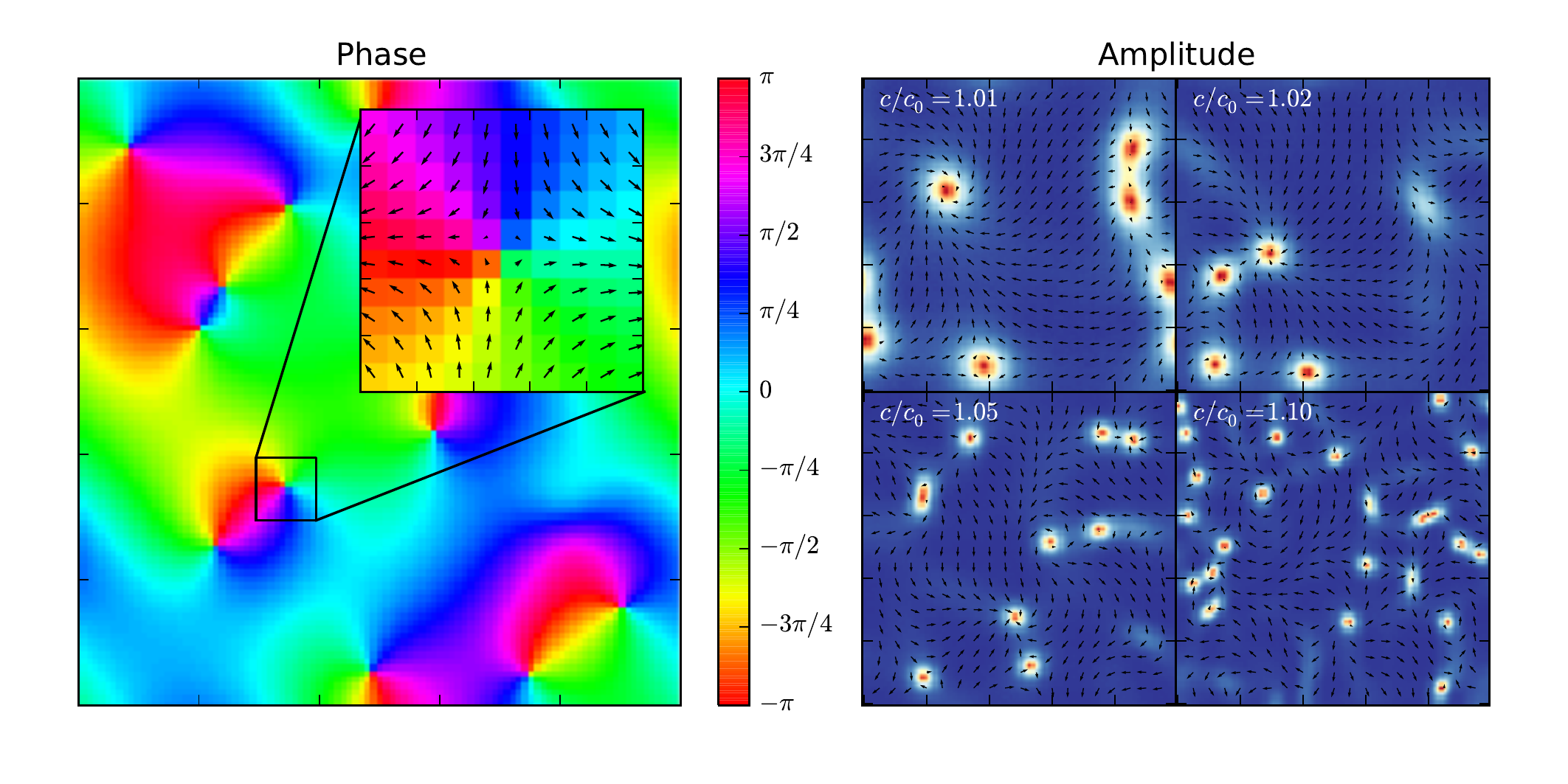}
\caption{Phase (left) and amplitude (right) for 2D XY model with vortices.}
\label{fig:09-f15}
\end{center}
\end{figure}

\subsubsection{Vortex Interactions}

A single vortex is a stable fixed point of the near-threshold equations of motion (\ref{eq:09-static-norm2}).  But if two vortices are placed together, the solution is no longer stable.  Far from the vortex cores $|r| \gtrsim \ell$, the field amplitude is constant and only its phase varies: $A(x, y) = a_{\rm sat} e^{i\phi(x, y)}$.  The equation of motion for $\phi(\bar{x}, \bar{y})$ is:
\beq
	\frac{\partial\phi}{\partial\bar{t}} = \frac{1}{2}\nabla^2 \phi
\eeq

One finds the vortex attraction in a manner analogous to Eq.~(\ref{eq:09-vw}) (for 1D domain-wall attraction).  For the solution $\phi_z(\bar{x}, \bar{y}) = \mbox{Im}[\log((\bar{x}-\bar{z}) + i\bar{y}) - \log((\bar{x}+\bar{z}) + i\bar{y})]$, which is parameterized by the separation $2z$, all perturbations decay rapidly except the translation modes $\partial\phi_z/\partial\bar{x}$, $\partial\phi_z/\partial\bar{y}$, and the attraction mode $\partial\phi_z/\partial \bar{z} = \bar{x}\,\partial\phi_z/\partial \bar{x}$.  The vortex attraction can be computed:
\beq
	\bar{v}_{\rm vort} = -\left[\int\frac{\partial\phi}{\partial\bar{z}}\frac{\partial\phi}{\partial\bar{z}}d\bar{x}d\bar{y}\right]^{-1}
		\int\frac{\partial\phi}{\partial\bar{t}}\frac{\partial\phi}{\partial\bar{z}}d\bar{x}d\bar{y} \label{eq:09-vvort}
\eeq
The first term is an inertial term.  One can compute it by noting that the ``inertia'' of two vortices is roughly twice that of a single vortex, and for a single vortex,
\beq
	\int{\frac{\partial\phi}{\partial\bar{z}}\frac{\partial\phi}{\partial\bar{z}}d\bar{x}d\bar{y}} = \frac{1}{2}\int{(\nabla\phi)^2 d\bar{x}d\bar{y}} = \pi \ln(R/r_0)
\eeq

This is infinite for a single vortex when $R \rightarrow \infty$, consistent with the well-known fact that individual vortices have infinite energy in the XY model.  For a vortex pair at $(z, -z)$ one can set $R \approx z$.  The denominator $r_0$ is set by the lattice size in the classical XY model, of the vortex size $\ell$ here.  The full inertial term will match up to a numerical factor: $A\pi\ln(z/r_0)$.

\begin{figure}[tbp]
\begin{center}
\includegraphics[width=1.00\textwidth]{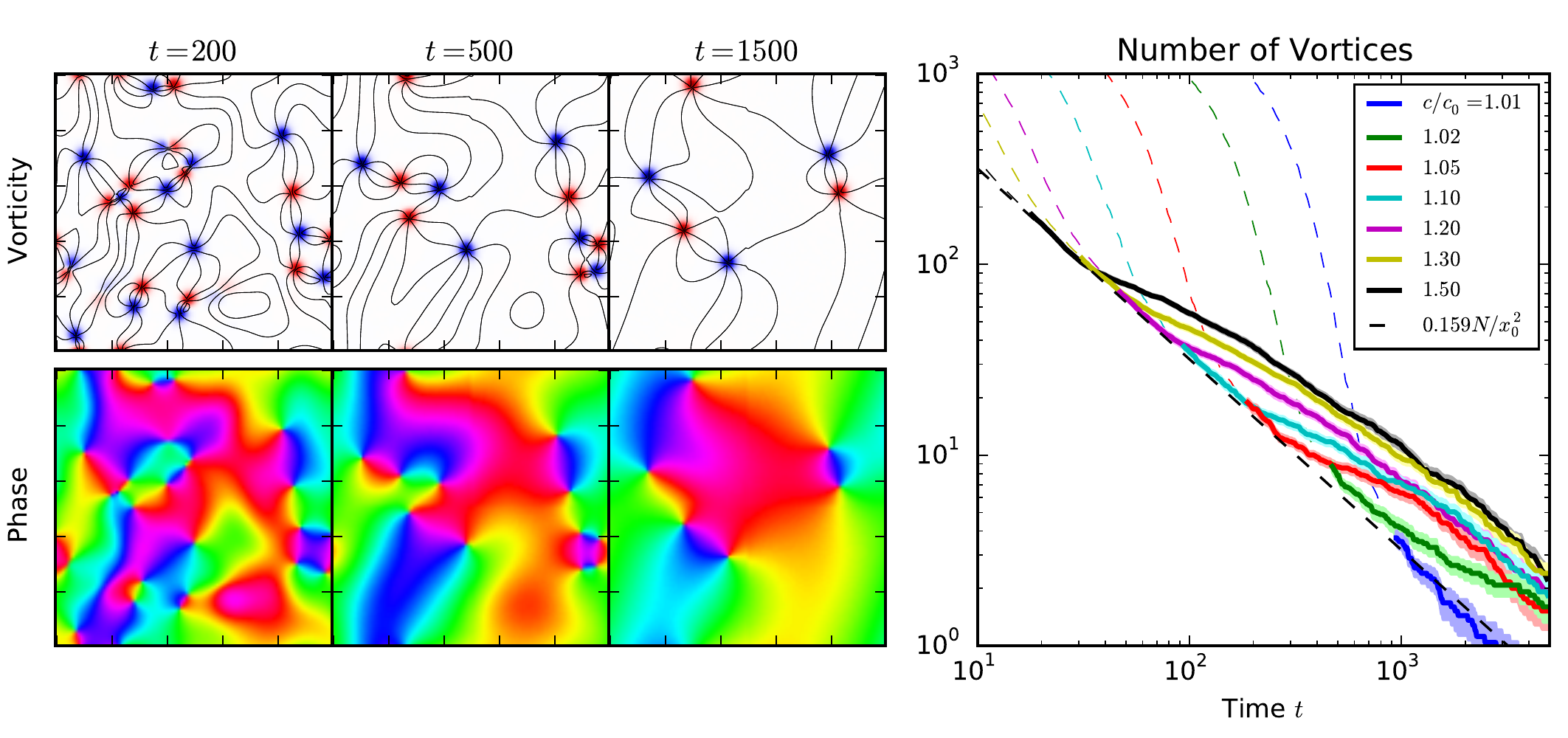}
\caption{Left: plots of vorticity and phase as a function of time (simulation used $c/c_0 = 1.1$).  Right: vortex number as a function of time and pump power ($t > T_{\rm sat}$ in bold).}
\label{fig:09-f16}
\end{center}
\end{figure}

The second term in (\ref{eq:09-vvort}) is a forcing term.  Since $\partial\phi/\partial\bar{t} = \tfrac{1}{2}\nabla^2\phi$, one can rewrite it in terms of a potential, which may be calculated by methods of complex analysis; see Kosterlitz \& Thouless\cite{Kosterlitz1973}:
\beq
	\int\frac{\partial\phi}{\partial\bar{t}}\frac{\partial\phi}{\partial\bar{z}}d\bar{x}d\bar{y} = 
		-\frac{1}{4}\frac{\d}{\d z}\int{(\nabla\phi)^2 d\bar{x}d\bar{y}} = -\frac{1}{4}\frac{\d}{\d\bar{z}}\left[4\pi \ln\frac{2\bar{z}}{r_0}\right] = -\frac{\pi}{\bar{z}}
\eeq
Thus, the attraction speed for a vortex pair at $(z, -z)$, and the lifetime for such a pair, is:
\beq
	\bar{v}_{\rm vort} \approx \frac{1}{A\bar{z}\ln (\bar{z}/r_0)},\ \ \ 
	\bar{T} = \frac{A}{2}\bar{z}^2\left(\log(\bar{z}/r_0) - \tfrac{1}{2}\right) \label{eq:vwxy}
\eeq
In real units, $T = \tau\bar{T}, z = \ell\bar{z}$, so a coefficient $\tau/\ell^2$ appears on the right-hand side.  This evaluates to:
\beq
	v_{\rm vort} = \frac{r^2(1-r^2)}{A} \frac{1}{z \ln(z/r_0\ell)},\ \ \ 
	T = \frac{A}{r^2(1-r^2)} z^2\left(\log(z/r_0\ell) - \tfrac{1}{2}\right) \label{eq:09-tvar}
\eeq
Unlike for domain walls, the vortex collision time scales only logarithmically with the pump, in that $\ell$ renormalizes the cutoff length $r_0$.  It also scales polynomially with $z$, suggesting that for an $L\times L$ lattice the system should reach the ground state (up to a winding number) in $O(L^2 \log L)$ time, similar to the $O(L^2)$ scaling found for 1D chains in Sec.~\ref{sec:xy1d}.  One can run 2D XY-machine simulations with two-vortex initial conditions; the vortex attraction roughly follows (\ref{eq:09-tvar}) with the parameters $A \approx 1.80$, $r_0 \approx 0.25$.

Figure \ref{fig:09-f16} illustrates the vortex interactions, albeit qualitatively.  For a complex field of arbitrary amplitude, one can define the vorticity as $\nabla a^* \times \nabla a$.  The winding number around a loop (for which the field has constant amplitude) equals the integral of the vorticity inside the loop.  This vorticity is plotted as a function of position and time, and the regions of nonzero vorticity correspond to regions where the phase wraps by $2\pi$.  Following the plots from left to right, one sees that vortices of opposite vorticity are attracted to each other and eventually annihilate, consistent with the vortex interaction picture sketched above.

The right plot shows the average number of vortices on a $100\times 100$ XY model for pump amplitudes ranging from $c/c_0 = 1.01$ to $1.50$.  In this plot, the ``number of vortices'' was defined as the number of unit cells in which the phase winds by $\pm 2\pi$.  Well below saturation, when the field amplitudes are random, the number of such ``vortices'' is very high.  By the end of the growth stage, the vortex count stabilizes at $0.159N/x_0^2$, consistent with Eq.~(\ref{eq:09-nvort}).  Thereafter the system enters the Kuramoto stage and its dynamics are driven by vortex-vortex interactions.  From (\ref{eq:09-tvar}), we expect that the number of vortices should scale as $N_v \sim T^{-1}$ up to a logarithmic term; this explains the near-linear falloff of all of the curves on the log-log scale.

\section{Conclusion}

Although the Ising problem is quite old, the OPO-based Ising solver is a new idea.  This paper presents the first comprehensive treatment of 1D and 2D ferromagnetic Ising and XY machines based on this mechanism.  The Ising machine differs from simulated and quantum annealing in that the ``spins'' are not bits or qubits, but rather optical states in an OPO.  The dynamics of this OPO network can be simulated using semiclassical equations derived from the truncated Wigner method.

Previous papers modeled the Ising machine as a network of coupled cavities and derived continuous-time equations of motion for the state\cite{Haribara2015,Utsunomiya2011,Wang2013}.  In the time-multiplexed picture (Fig.~\ref{fig:09-f1}), that approach is only valid when the cavity has high finesse ($G_0 \approx 1$) and the round-trip coupling between pulses is weak ($r \ll 1$).  Thus, the coupled-cavity model is not accurate for high-gain systems like those at RIKEN\cite{KentaThesis,Takata2016}, NTT\cite{Inagaki2016} and Stanford\cite{McMahon2016}.  On the other hand, high-gain systems are advantageous because they are faster and more resilient to experimental noise and loss.

In this paper, we derive a more general approach which holds for cavities of arbitrary finesse and coupling.  The truncated Wigner picture is used and the state is described by semiclassical pulse amplitudes $a_i(t)$, where $i$ is the pulse index and $t$ is the {\it discrete} time (round-trip number).  This state satisfies a set of {\it difference equations} (Eqs.~(\ref{eq:feom}, \ref{eq:09-aout}, \ref{eq:09-ai-delay2})) that relate $a_i(t+1)$ to $a_i(t)$.  These equations reduce to the continuous-time equations in the high-finesse limit $G_0 \rightarrow 1$, $r \rightarrow 0$.

Both 1D and 2D Ising chains were simulated using this model.  The dynamics of the Ising machine can be broken into two stages: a {\it growth stage} (Sec.~\ref{sec:09-growth}) in which the field amplitudes are far below saturation, and a {\it saturation stage} (Sec.~\ref{sec:09-crst}), by which most of the OPO amplitudes have saturated.  During the growth stage, the OPO amplitudes start from random values and grow linearly, with longer-wavelength Fourier modes growing the fastest.  This induces correlations between nearby OPOs, forming ferromagnetic domains after saturation.  During the saturation stage, these domains evolve, and the attraction of nearby domain walls causes smaller domains to annihilate.

We used this model to compute basic statistical quantities in 1D: the correlation function $R(x)$, correlation length $x_0$, defect density $n_d$, domain length distribution $P(\ell)$, and success probability $P_s$.  In the Ising machine, these all depend on the time to saturation $T$ (which is a function of pump rate) and the coupling mirror parameters $r, t$; for the thermal model they are functions of the coupling $J$ and effective temperature $1/\beta$.  Experimental data from Inagaki et al.\cite{Inagaki2016} match closely with our numerical predictions.

The dynamics depend strongly on dimension.  For the 1D chain, the domain lifetime scales exponentially with domain size, so one can say that after the growth stage, the domain structure ``freezes out'', and will not relax to the ferromagnetic ground state unless one waits an exponentially long time.  Conversely, in the 2D case this lifetime scales as the size squared, since domain walls are curved and always move towards their center of curvature (Sec.~\ref{sec:09-2dlat}).  Thus, long-range order is established in $O(L^2)$ time for an $L \times L$ lattice, and all domain walls are eventually destroyed.

Ising simulations for frustrated 1D and 2D systems were also studied.  In this case, the Fourier modes with maximum gain have nonzero $k$, giving rise to periodic order in the final state (Fig.~\ref{fig:09-f11}).  In 2D, one finds two competing phases of periodic order: up- and down-diagonal stripes, which compete with each other, analogous to the competition between up- and down-states in the Ising model (Fig.~\ref{fig:09-f12}).

We also studied a related OPO network, the coherent XY machine.  This device uses a network of {\it non}-degenerate OPOs to find the ground state of the XY potential.  If the XY machine is based on pulses in a high-gain cavity with strong couplings (Fig.~\ref{fig:09-f13}), one obtains a similar set of difference equations, this time for both signal and idler fields.  As before, if we take the limit $G_0 \rightarrow 1$, $r \rightarrow 0$, this reduces to the continuous-time coupled-cavity model studied elsewhere.

For 1D XY systems, the only possible topological defect is the winding number.  In $O(N^2)$ time, the system always relaxes to a state with constant winding.  Before this ``smoothing out'', winding can be treated as a random walk per unit length, and the winding number has a Gaussian with standard deviation that goes as $N^{1/2}$.  For 2D systems, vortex defects form, analogous to the BKT transition\cite{Kosterlitz1973}.  In contrast to BKT, vortices in the XY machine are formed through the OPO growth / saturation process, not thermally; thus their distribution is very different.

It is hoped that our results for these simple models will shed insight into Ising machines more generally.  From the results above, a few things stand out:

\begin{enumerate}
\item As an ``algorithm'', the Ising machine is behaving like a convex relaxation technique.  Dividing the dynamics into growth and saturation stages makes this more obvious.  During the growth stage, the eigenmodes of the coupling matrix $C_{ij}$ grow at different rates and the machine tends toward the dominant eigenmode.  This is solving the maximum eigenvalue problem, which has a single local minimum and is solvable in polynomial time (although it is not technically convex).  However, this eigenmode may not be a valid Ising state, so during the saturation stage, the system relaxes into a valid state as the pulse amplitudes saturate.
\item Unlike simulated annealing, randomness does not appear to play a major role in this algorithm.  While random noise seeds the initial state, most of the subsequent dynamics is deterministic because the field amplitudes are far above the quantum level.  When the system reaches a local minimum in the saturation stage, it is unable to ``tunnel'' out (in either a classical or quantum sense) because the photon number is so high.
\item Even ``trivial'' problems can have long-lived metastable states (e.g.\ domain walls) or local minima (winding numbers).  The current machine does not have a way to escape these minima, since the noise is so small compared to the coherent amplitude at saturation.  However, it is equally worth mentioning that simulated annealing is not very efficient on the 1D chain, requiring at least $O(N^2)$ time to converge.  For the Ising machine, the convergence time is also $O(N^2)$, if this time is spent during the growth stage.
\end{enumerate}

While so far only the 1D chain has studied experimentally, by adding extra delay lines, it is straightforward to extend current work to the 2D and frustrated cases.  Moreover, the groups at Stanford\cite{McMahon2016} and NTT\cite{Inagaki2016b} are working towards machines with ``all-to-all'' connectivity via injection and measurement feedback\cite{Haribara2016}.  The measurement-feedback theory is probably a straightforward extension of this work, with additional stochastic terms for detector, ADC/DAC and injection noise.  Beyond the scope of this work, the measurement-feedback approach is promising because it can handle arbitrary spin networks, not just the 1D and 2D lattices of this paper.

\section*{Acknowledgements}

The authors thank Shoko Utsunomiya and Shuhei Tamate for useful discussions.  Ryan Hamerly is funded by a seed grant from the Precourt Institute for Energy at Stanford University.  This research is supported by the Impulsing Paradigm Change through Disruptive Technologies (ImPACT) Program of the Council of Science, Technology and Innovation (Cabinet Office, Government of Japan).

\bibliography{NoteRefs}{}
\bibliographystyle{plain}

\end{document}